\documentclass{aa}
\usepackage{txfonts}
\usepackage{graphicx}
\usepackage{multicol}
\usepackage{amssymb}
\usepackage{color}
\usepackage{tabularx}
\usepackage{amsmath}
\begin{document}

\title{Core-collapse supernova progenitor constraints using the spatial distributions of massive stars in local galaxies}
\author{T. Kangas\inst{1}, L. Portinari\inst{1}, S. Mattila\inst{1,2,3}, M. Fraser\inst{3}, E. Kankare\inst{4}, R. G. Izzard\inst{3}, P. James\inst{5}, \\ C. Gonz\'{a}lez-Fern\'{a}ndez\inst{3}, J. R. Maund\inst{6,7} and A. Thompson\inst{4}
          }

\institute{Tuorla Observatory, Department of Physics and Astronomy, University of Turku, V\"{a}is\"{a}l\"{a}ntie 20, FI-21500 Piikki\"{o}, Finland\\
              \email{tjakan@utu.fi}
         \and
Finnish Centre for Astronomy with ESO (FINCA), University of Turku, V\"{a}is\"{a}l\"{a}ntie 20, FI-21500 Piikki\"{o}, Finland
         \and
Institute of Astronomy (IoA), University of Cambridge, Madingley Road, Cambridge, CB3 0HA United Kingdom
         \and
Astrophysics Research Centre, School of Mathematics and Physics, Queen's University Belfast, BT7 1NN, UK
         \and
Astrophysics Research Institute, Liverpool John Moores University, IC2, Liverpool Science Park, 146 Brownlow Hill, Liverpool, L3 5RF, UK
 	 \and
The Department of Physics and Astronomy, The University of Sheffield, Hicks Building, Hounsfield Road, Sheffield, S3 7RH, UK
	 \and
Royal Society Research Fellow
             }

\date{Accepted for publication in Astronomy \& Astrophysics}

  \abstract
{We study the spatial correlations between the H\,$\alpha$ emission and different types of massive stars in two local galaxies, the Large Magellanic Cloud (LMC) and Messier 33. We compare these to correlations derived for core-collapse supernovae (CCSNe) in the literature to connect CCSNe of different types with the initial masses of their progenitors and to test the validity of progenitor mass estimates which use the pixel statistics method. We obtain samples of evolved massive stars in both galaxies from catalogues with good spatial coverage and/or completeness, and combine them with coordinates of main-sequence stars in the LMC from the SIMBAD database. We calculate the spatial correlation of stars of different classes and spectral types with H\,$\alpha$ emission. We also investigate the effects of distance, noise and positional errors on the pixel statistics method. A higher correlation with H\,$\alpha$ emission is found to correspond to a shorter stellar lifespan, and we conclude that the method can be used as an indicator of the ages, and therefore initial masses, of SN progenitors. We find that the spatial distributions of type II-P SNe and red supergiants of appropriate initial mass ($\gtrsim$9 $M_{\odot}$) are consistent with each other. We also find the distributions of type Ic SNe and WN stars with initial masses $\gtrsim$20 $M_{\odot}$ consistent, while supergiants with initial masses around 15 $M_{\odot}$ are a better match for type IIb and II-L SNe. The type Ib distribution corresponds to the same stellar types as type II-P, which suggests an origin in interacting binaries. On the other hand, we find that luminous blue variable stars show a much stronger correlation with H\,$\alpha$ emission than do type IIn SNe.}

\authorrunning{Kangas et al.}
\titlerunning{Supernova progenitor constraints using the spatial distributions of massive stars in local galaxies}

   \maketitle
%

\section{Introduction}

Our understanding of the late evolutionary stages and final fates of massive stars (with masses above $\sim$8 $M_{\odot}$), despite decades of theoretical and observational work devoted to it, remains incomplete. Generally, these stars are expected to end their lives as core-collapse supernovae (CCSNe) after evolving through a series of stages of nuclear burning until they have built up an iron core. The type of the resulting SN explosion is expected to depend on the initial mass of the progenitor star, along with an as yet unclear contribution from a number of other factors such as the metallicity, rotation and multiplicity of the star \citep[e.g.][and references therein]{smartt09, smith11a, georgy12, eldridge13}. 

CCSNe are divided into two main types, hydrogen-poor type Ib/c SNe (type Ic also being helium-poor) and hydrogen-rich type II SNe \citep[][]{sntypes}. Type Ib and Ic SNe are believed to have progenitors that have lost their hydrogen and/or helium envelopes, while the progenitors of type II SNe still retain at least a part of their envelope, resulting in hydrogen lines in their spectra \citep[e.g.][]{smartt09}. Type II SNe are also divided into subtypes. The red supergiant (RSG) progenitors of type II-P (`plateau') SNe still retain massive hydrogen envelopes that power the distinctive plateaus in their light curves \citep{gin71}, while type II-L (`linear') progenitors are believed to have lost a significant part of their hydrogen envelope and type IIb progenitors almost all of it. Although originally considered two distinct populations of events \citep[e.g.][]{barbon79}, types II-P and II-L are now more commonly thought to occupy a continuum of different levels of mass loss primarily influenced by the initial mass of the progenitor \citep[e.g.][]{a14,sanders15,gonzalez}. Type IIn (`narrow lines') SNe are characterized by the presence of a dense circumstellar medium (CSM) at the time of explosion, resulting in strong interaction between the SN ejecta and the CSM \citep[e.g.][]{schlegel90, 1988z}. Non-terminal outbursts of massive stars in external galaxies, analogous to the Great Eruption of $\eta$ Car, have sometimes been mistaken for true SNe, and are called SN impostors \citep[e.g.][]{vandyk00, pastorello13, kankare15}. 

For type Ib and Ic SNe, both very massive single Wolf-Rayet (WR) stars \citep[e.g.][]{maeder82} and lower-mass stars stripped through interaction with a binary companion \citep[e.g.][]{pjh92} have been suggested as progenitors, and \citet{eldridge13} suggested a mix of both progenitor channels. Recently, \citet{smith14} challenged the established picture of single-star mass loss; namely, that mass loss through conventional line-driven winds should not be sufficient to create a type Ib/c SN, and most of their progenitors should be stripped through binary interaction. \citet{lyman16} found the ejecta masses of type Ib/c and IIb SNe inconsistent with very massive stars, also indicating that interacting binaries are the dominant progenitor channel. The effects of multiplicity on massive star evolution, however, are not well understood either, despite its tremendous importance \citep[e.g.][]{sana12,sana13}. Properly taking into account Roche-lobe overflow, mergers and other features of binaries is thus one of the most important challenges in understanding the evolution of massive stars.

Empirical studies of the progenitors of different types of SNe can help distinguish between different evolutionary schemes. Direct detections of type II-P SN progenitors in pre-explosion high-resolution images \citep[e.g.][]{mattila08, vandyk+12, maund13, fraser16} have already helped to establish them as RSGs with initial masses of $\sim$8.5 to $\sim$16.5 $M_{\odot}$ \citep{smartt09, smartt15}. 

The detected type IIb progenitors are consistent with interacting binary systems \citep[e.g.][]{aldering94, maund04, vandyk11, folatelli14, fox14, maeda14, folatelli15}, and a WR progenitor was ruled out for SN 2008ax \citep{folatelli15} although the proposed binary companion of the SN 2011dh progenitor was disputed by \citet{maund15a}. \citet{maeda15} found that the more extended progenitors may still be undergoing binary interaction, while for the less extended ones this phase would be over. The detected type IIb progenitors have been identified as yellow or blue supergiants (YSGs or BSGs, respectively) with initial masses between 13 and 19 $M_{\odot}$ \citep[][]{vandyk11, maund11, vandyk14, folatelli15}.

Among other SN types, however, direct progenitor detections are scarce. A few type IIn events have been connected to a luminous blue variable (LBV) progenitor \citep[e.g.][]{galyam09, smith11b, fransson14}, along with events such as SN 2009ip and SN 1961V where the disappearance of the progenitor has not been firmly ascertained \citep[e.g.][]{vandyk12, fraser15}. The (probably) only detected progenitor of a type II-L SN, that of SN 2009hd, was likely a high-mass RSG \citep[][]{2009hd}. A possible progenitor for another type II-L, SN 2009kr \citep{fraser10}, was shown by \citet{maund15b} to probably be a small cluster instead of a single star. 

No type Ic progenitors have been detected so far, possibly because of their faintness in the optical bands \citep{yoon12}. The only detected type Ib progenitor, that of iPTF13bvn \citep{cao13}, was suggested to be an initially 10 -- 20 $M_{\odot}$ star in an interacting binary by \citet{eldridge15}. More recently \citet{em16} suggested an initially 10 -- 12 $M_{\odot}$ star that evolved into a helium giant; \citet{folatelli16} also favored a binary progenitor, although none of the models they examined proved entirely satisfactory. \citet{vandyk16} excluded a $\gtrsim$10 $M_{\odot}$ companion to the progenitor of the type Ic SN 1994I, which earlier had been considered the result of binary interaction \citep{nomoto94}. \citet{heikkila16} used pre-explosion X-ray observations to investigate the possibility of high-mass X-ray binaries (in which one component is a neutron star or a stellar-mass black hole) being the progenitors of some type Ib, Ic or IIb SNe. Such progenitors were found to be rare.

In lieu of direct detections, SN progenitors can also be studied in various indirect ways. \citet{hakobyan09,hakobyan16}, \citet{h10} and \citet{h12}, for example, examined the distances of different types of SNe from the nuclei of their host galaxies, and found type Ib/c SNe to be more centrally concentrated than type II. \citet{kk12} used the colours of SN environments to suggest a higher initial mass for type Ib/c progenitors than those of type II. \citet{hanin_a} used integral field spectroscopy to infer a higher environmental metallicity and initial mass for type Ic progenitors than type Ib, while \citet{hanin_b} found some type II progenitors to have initial masses comparable to type Ib/c progenitors. \citet{leloudas10} connected type Ib/c SNe and gamma-ray bursts to WR stars using the relative \emph{B}- and \emph{K}-band brightness of their environments. Spectral synthesis model fitting to spectra of SNe at the nebular phase has been used by \citet{mazzali10} and \citet{jerkstrand15} to suggest an initially 15 $M_{\odot}$ progenitor for type Ic SN 2007gr and a type IIb progenitor mass range of 12 -- 16 $M_{\odot}$, respectively. \citet{tomasella13} used hydrodynamical modeling to constrain the ejecta mass of a type II-P event, SN 2012A, to 12.5 $M_{\odot}$. \citet{do14} suggested an envelope mass of $\sim20$ $M_{\odot}$ for SN 2012aw, another type II-P event, using similar methods, challenging the $\sim$8.5 to $\sim$16.5 $M_{\odot}$ initial mass range.

Because ionizing radiation from young massive stars is responsible for creating H~{\sc ii} regions \citep{kennicutt98}, correlations between different SN types and the H\,$\alpha$ emission from their host galaxies can also be used to statistically study their progenitors. \citet{ja06} -- hereafter JA06 -- used a method called pixel statistics to study this correlation, and this method was used again with larger samples by \citet{aj08}~and \citet{a12} -- hereafter AJ08 and A12, respectively. These results indicated a higher average initial progenitor mass for type Ic SNe than types Ib, II-P or IIn. \citet{h14} -- hereafter H14 -- applied the same method to interacting transients, that is type IIn SNe and SN impostors. \citet{k13} -- hereafter K13 -- used both the pixel statistics method and the distances to host galaxy nuclei to study SNe in strongly star-forming galaxies specifically, and also found the A12 mass sequence (with a stronger correlation between type Ic and H\,$\alpha$ emission than in normal galaxies) and an increased centralization for type Ib/c. The results from H\,$\alpha$ images were also supported with near ultraviolet (NUV) pixel statistics by both A12 and K13. However, the pixel statistics method has so far remained qualitative. \citet{crowther13} argued that it can only provide weak constraints because the lifetimes of the giant H~{\sc ii} regions probed by A12 are longer than the lifetimes of some CCSN progenitors. The \citet{crowther13} method of measuring distances from the SNe to the nearest H~{\sc ii} regions, however, corroborated the A12 result. Indirect methods are also affected by biases such as the Malmquist bias when it comes to constructing a sample of SNe.

\begin{figure*}
\centering
\begin{minipage}{150mm}
\includegraphics[width=\columnwidth]{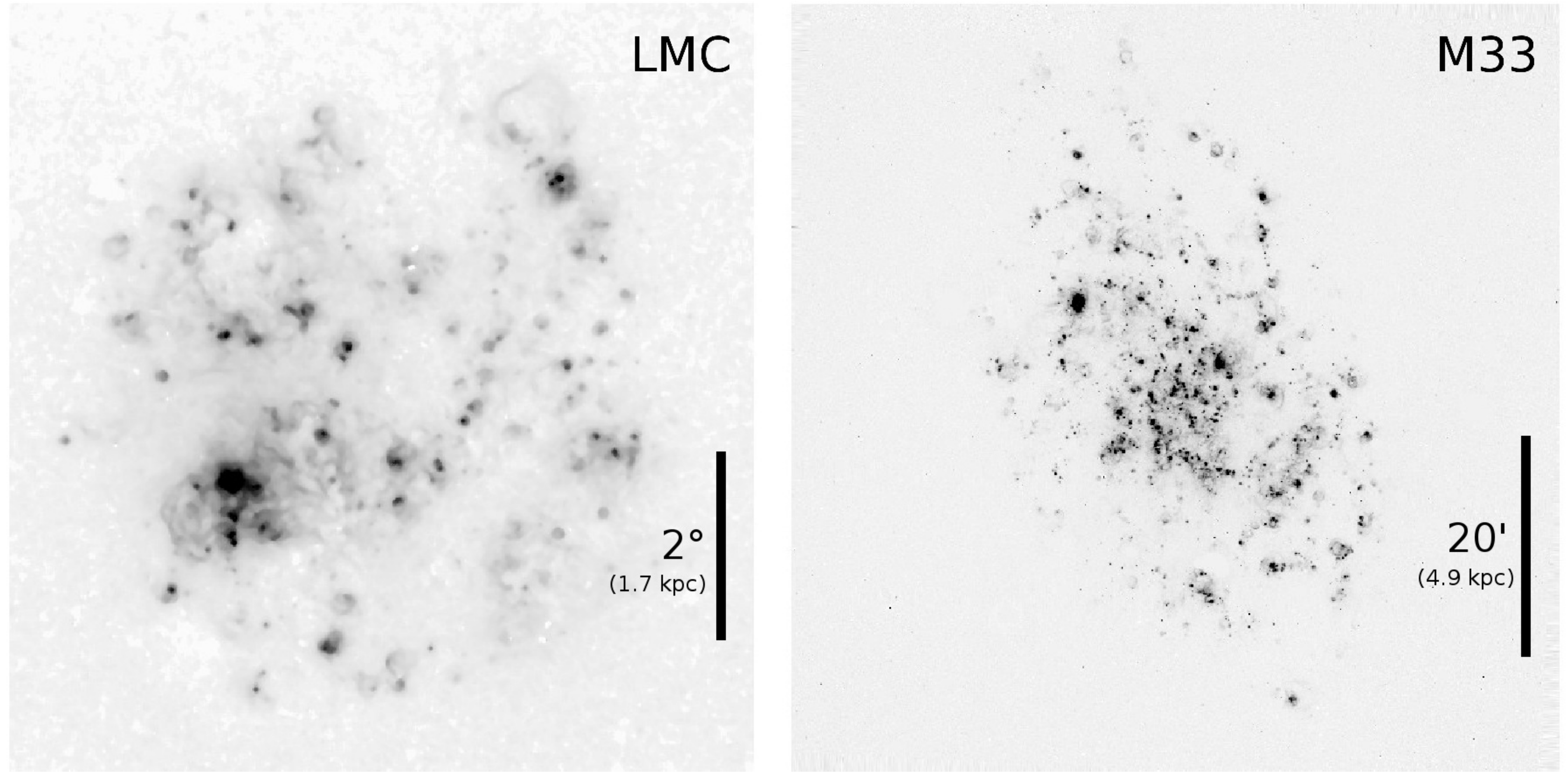}
\caption{Original continuum-subtracted H\,$\alpha$ intensity maps used in this study: the LMC on the left \citep[from the SHASSA survey;][]{lmcpic} and M33 on the right \citep[][]{m33pic}, cropped in the case of the LMC but otherwise unaltered. North is up and east is left. The black scale bar corresponds to 2 deg ($\sim$1.7 kpc) in the LMC and 20 arcmin ($\sim$4.9 kpc) in M33.}
\label{fig:ha_pics}
\end{minipage}
\end{figure*}

The results of A12 showed a weaker correlation with H\,$\alpha$ emission for type IIn than other CCSNe, indicating a lower initial progenitor mass. A12 argued that while the diversity within the subtype allows for some progenitors to be LBVs, most IIn progenitors are instead lower-mass stars, such as asymptotic giant branch (AGB) stars exploding as electron-capture SNe. To reconcile this with the identified massive LBV progenitors of type IIn SNe, \citet{smith15} examined the LBVs in the Large Magellanic Cloud (LMC) and found them to be relatively isolated from O-type stars, which was argued to at least partially explain the weak correlation with H\,$\alpha$ emission for the SNe. However, \citet{antismith} pointed out that this LBV sample included both classical and lower-luminosity post-RSG LBVs along with some unrelated stars, and that the classical LBVs are in fact closely associated with the O stars. A subset of ambiguous type IIn/Ia-CSM SNe also exists. It is unclear whether these events are thermonuclear \citep[][]{iacsm1} or core-collapse SNe \citep[][]{iacsm2}, which further complicates the already diverse type IIn. A fraction of type Ia SNe disguised as type IIn may partially explain the weak correlation with H\,$\alpha$ emission.

In this paper, we use H\,$\alpha$ intensity maps of two nearby galaxies, the LMC and Messier 33 (M33), to study the correlation between massive stars and star-forming regions in a way similar to A12 and K13. Nearby galaxies like these provide a unique opportunity for this study, as they are close enough for individual bright stars to be resolved and, unlike the Milky Way, we can see practically all of the H\,$\alpha$-emitting regions in them from the outside. Two very different galaxies are included in order to check the consistency of the results. We then compare these results to those of different CCSN types in other galaxies in order to test the validity of the pixel statistics method and attempt to derive more quantitative constraints for CCSN progenitors. We describe our H\,$\alpha$ images and the catalogues we use for stellar coordinates in Sect. 2. In Sect. 3 we describe our analysis methods. The results of the analysis are presented in Sect. 4, with an investigation of possible systematic effects in Sect. 5. Our findings are discussed in Sect. 6. Finally, in Sect. 7 we present our conclusions.


\section{Data}

Here we describe the H\,$\alpha$ images, catalogues of stellar coordinates and parameters, and SN samples used in the pixel statistics analysis.

\subsection{H\,$\alpha$ images}

We use two local galaxies to study the correlations between stars and H~{\sc ii} regions: the LMC, an SB(s)m-type dwarf galaxy at a well-established distance of $\sim$50 kpc, and M33, an SA(s)cd-type spiral galaxy at a distance between $\sim$700 and $\sim$900 kpc \citep[we adopt the most recent Cepheid distance, 839 kpc, from][]{m33dist}. The H\,$\alpha$ images of these galaxies are shown in Fig.~\ref{fig:ha_pics}, with no alterations except cropping in the case of the LMC. 

The continuum-subtracted H\,$\alpha$ image of the LMC was obtained from the Southern Hemisphere All-Sky Survey Atlas\footnote{http://amundsen.swarthmore.edu/} \citep[SHASSA; ][]{lmcpic}. The observations were performed using the Swarthmore robotic camera at the Cerro Tololo Inter-American Observatory (CTIO) in Chile. Details on the observations and reduction procedures are in the survey paper by \citet{lmcpic}. The uncropped image covers an area of 11.8 deg $\times$ 12.4 deg, which is cropped for our analysis into a size of 7.9 deg $\times$ 8.2 deg to only show the areas with H\,$\alpha$ emission. The pixel scale in the image is 47.64 arcsec pix$^{-1}$, corresponding to 11.6 pc pix$^{-1}$ at the distance of 50 kpc. The image has a spatial resolution (full width half maximum, FWHM) of 4.0 arcmin because of a median filter smoothing algorithm used in the survey for easier removal of foreground stars. At 50 kpc this corresponds to a resolution of 58.2 pc, which is sufficient for this study.

The continuum-subtracted H\,$\alpha$ image of M33 was obtained from the NASA/IPAC Extragalactic Database (NED)\footnote{http://ned.ipac.caltech.edu/}. This mosaic was originally made by \citet{m33pic} using images from the 0.6-m Burrell-Schmidt telescope at Kitt Peak National Observatory, taken in 1995. The image on NED covers an area of 1.15 deg $\times$ 1.15 deg. The image has a 2.03 arcsec pix$^{-1}$ scale and a spatial resolution (FWHM) of 3.0 arcsec, corresponding to 8.3 pc pix$^{-1}$ and 12.2 pc, respectively, at the distance of 839 kpc. For details on the observations and reductions, see \citet{m33pic}.

\subsection{Stellar samples}

\begin{figure}
\centering
\includegraphics[width=\columnwidth]{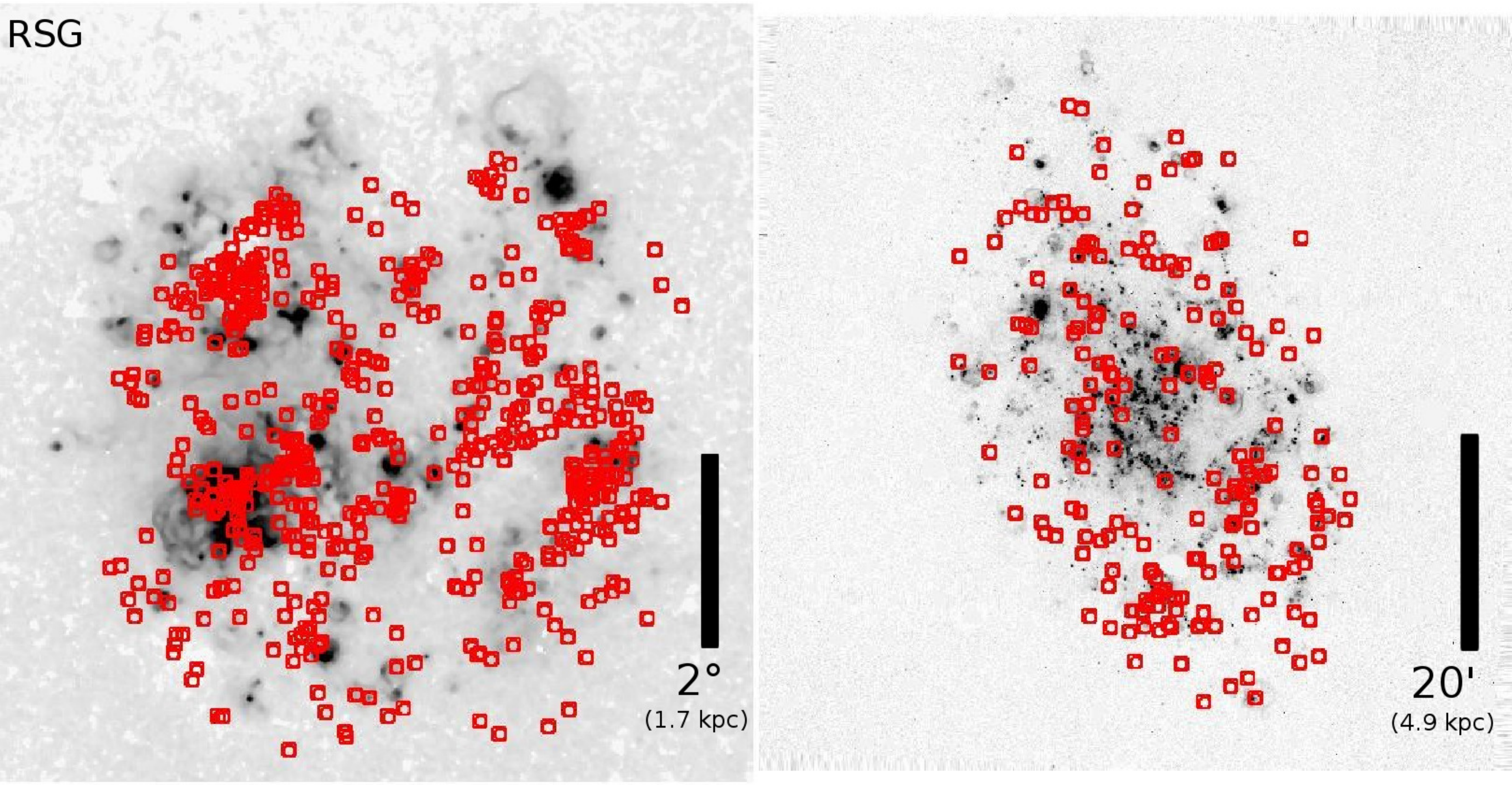}
\includegraphics[width=\columnwidth]{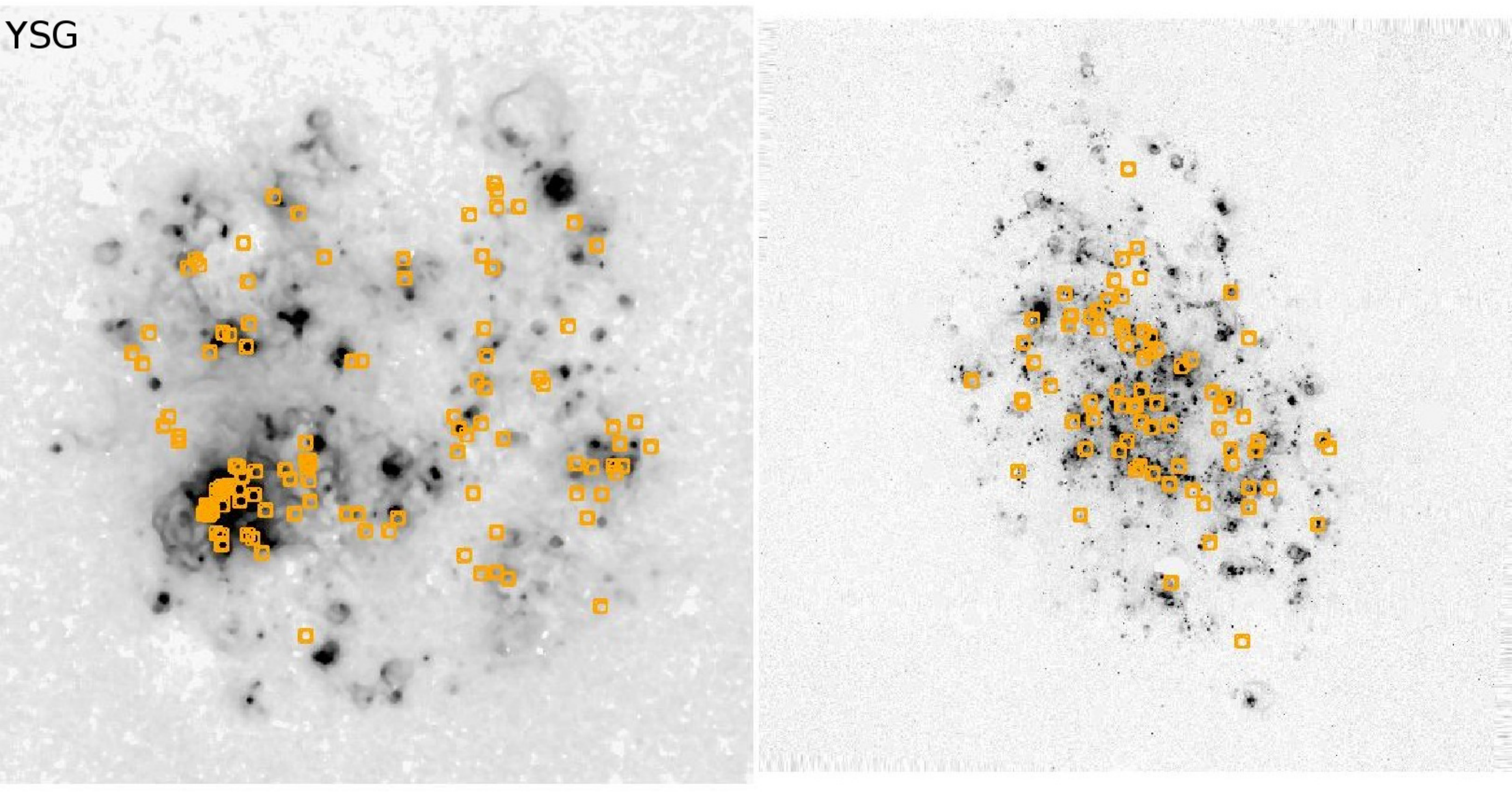}
\includegraphics[width=\columnwidth]{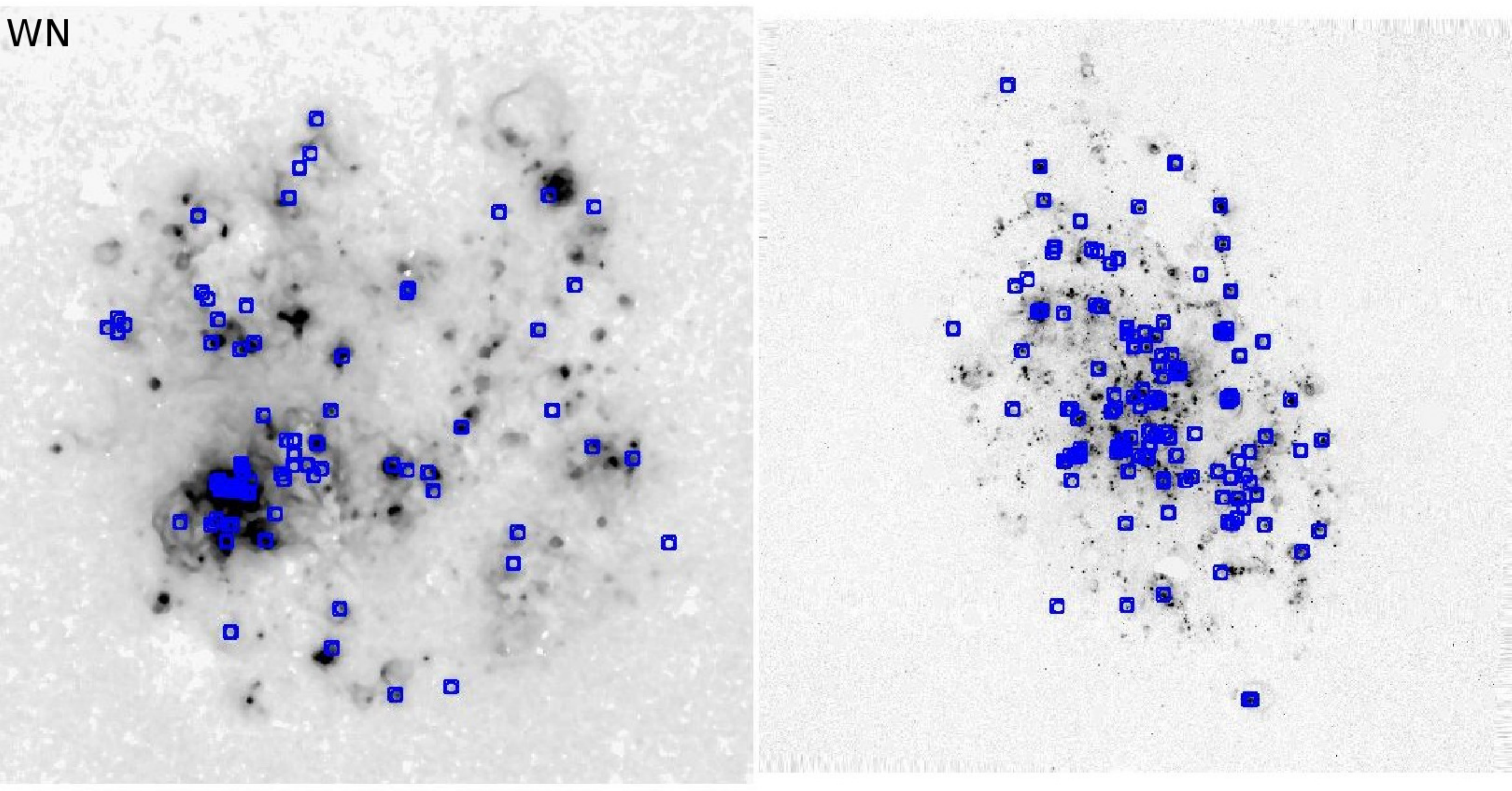}
\includegraphics[width=\columnwidth]{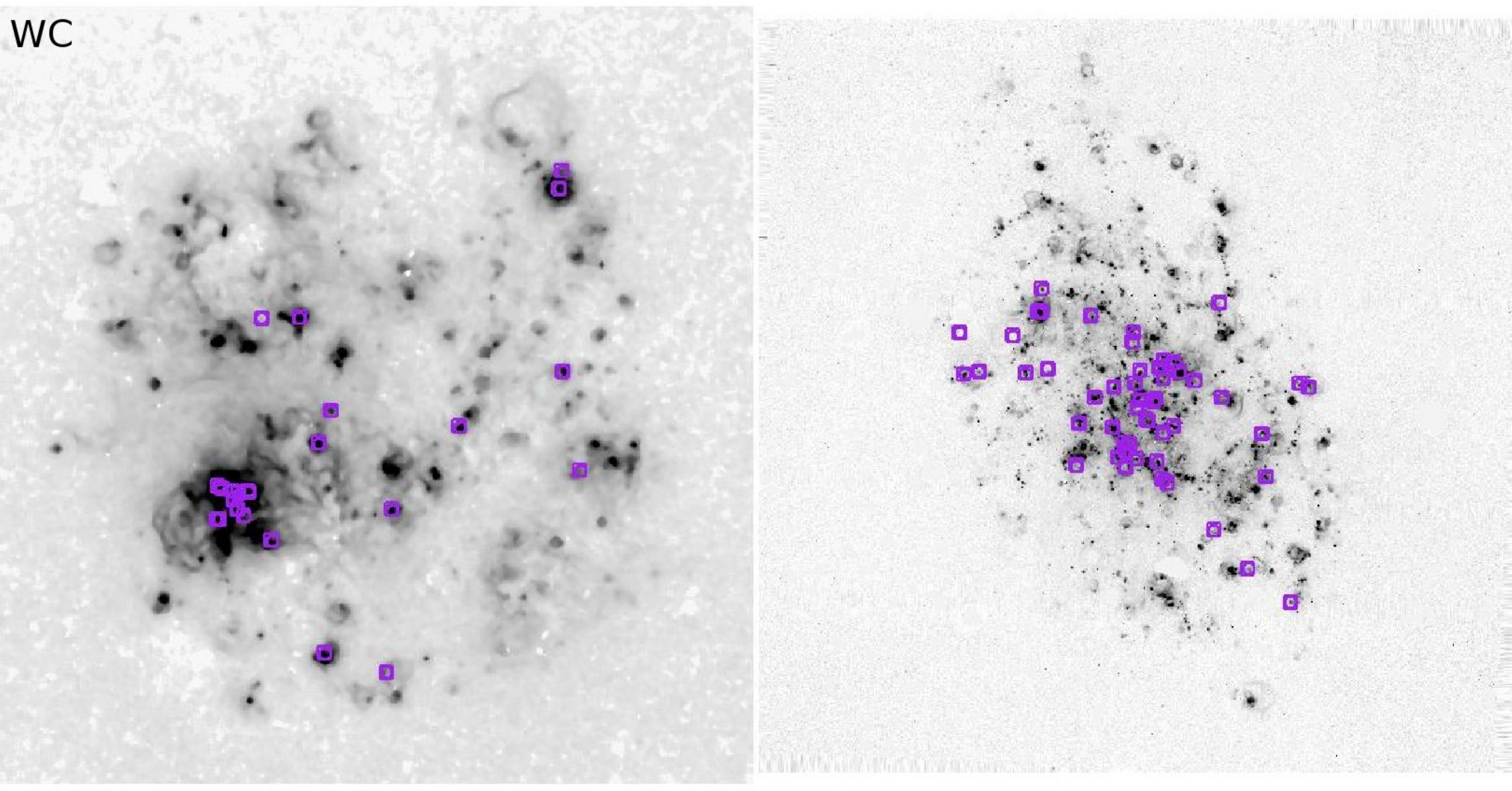}
\caption{H\,$\alpha$ images of the LMC (left) and M33 (right), with our RSG, YSG, WN and WC samples overplotted in order to illustrate the spatial coverage of their respective catalogues. The black scale bar corresponds to 2 deg ($\sim$1.7 kpc) in the LMC and 20 arcmin ($\sim$4.9 kpc) in M33. }
\label{fig:samples}
\end{figure}
\begin{figure}
\centering
\includegraphics[width=0.5\columnwidth]{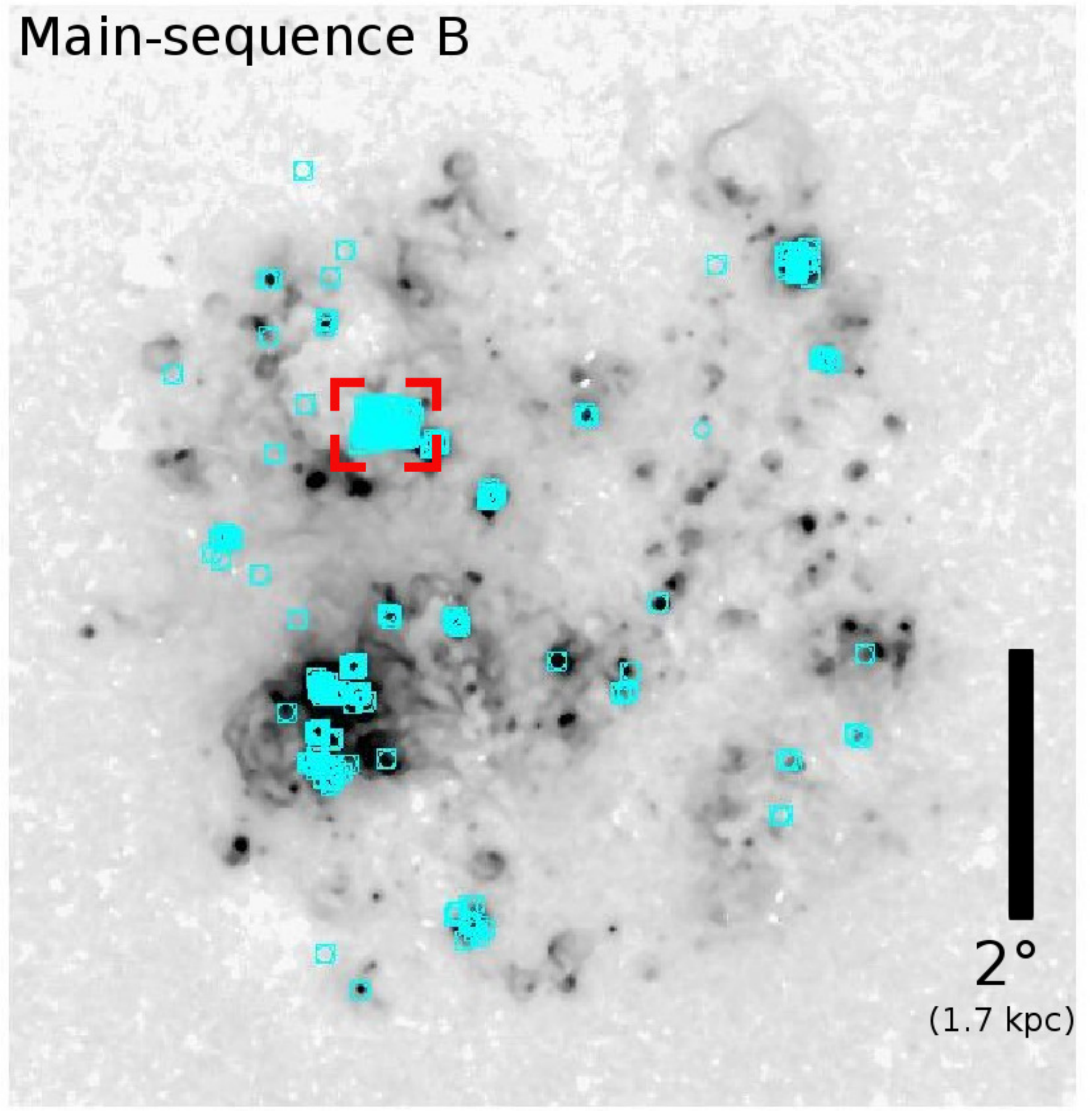}\includegraphics[width=0.5\columnwidth]{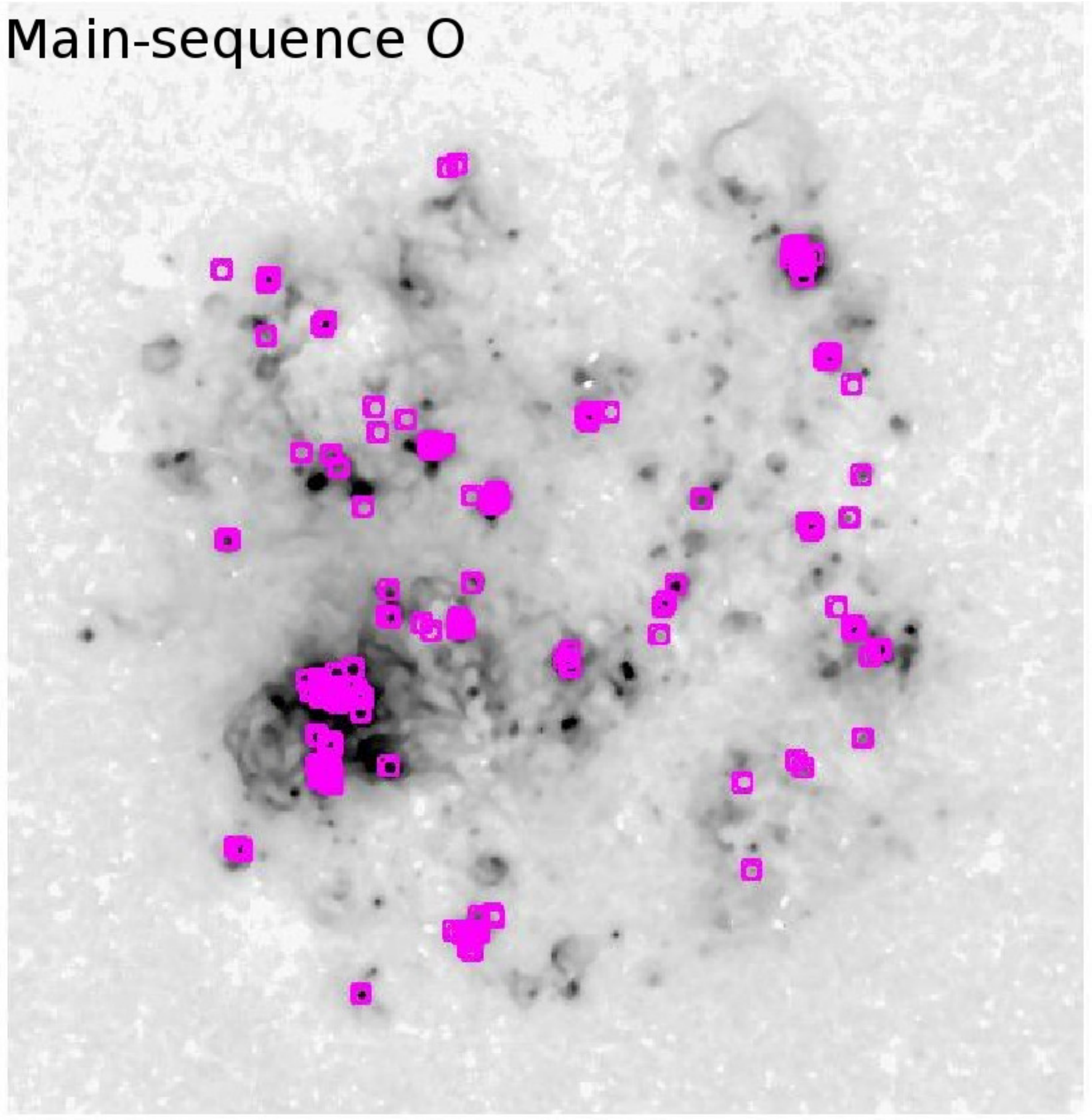}
\includegraphics[width=0.5\columnwidth]{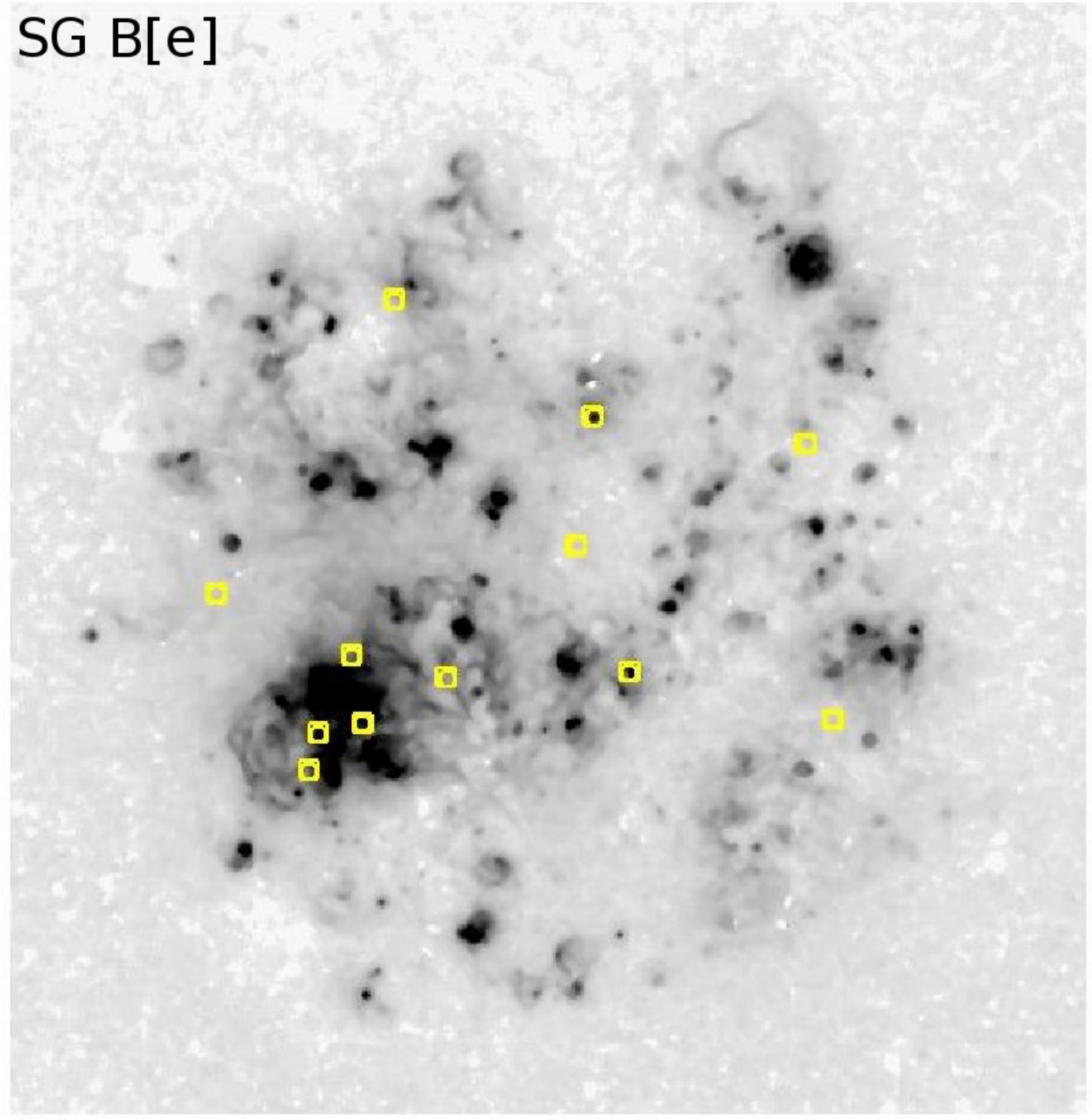}\includegraphics[width=0.5\columnwidth]{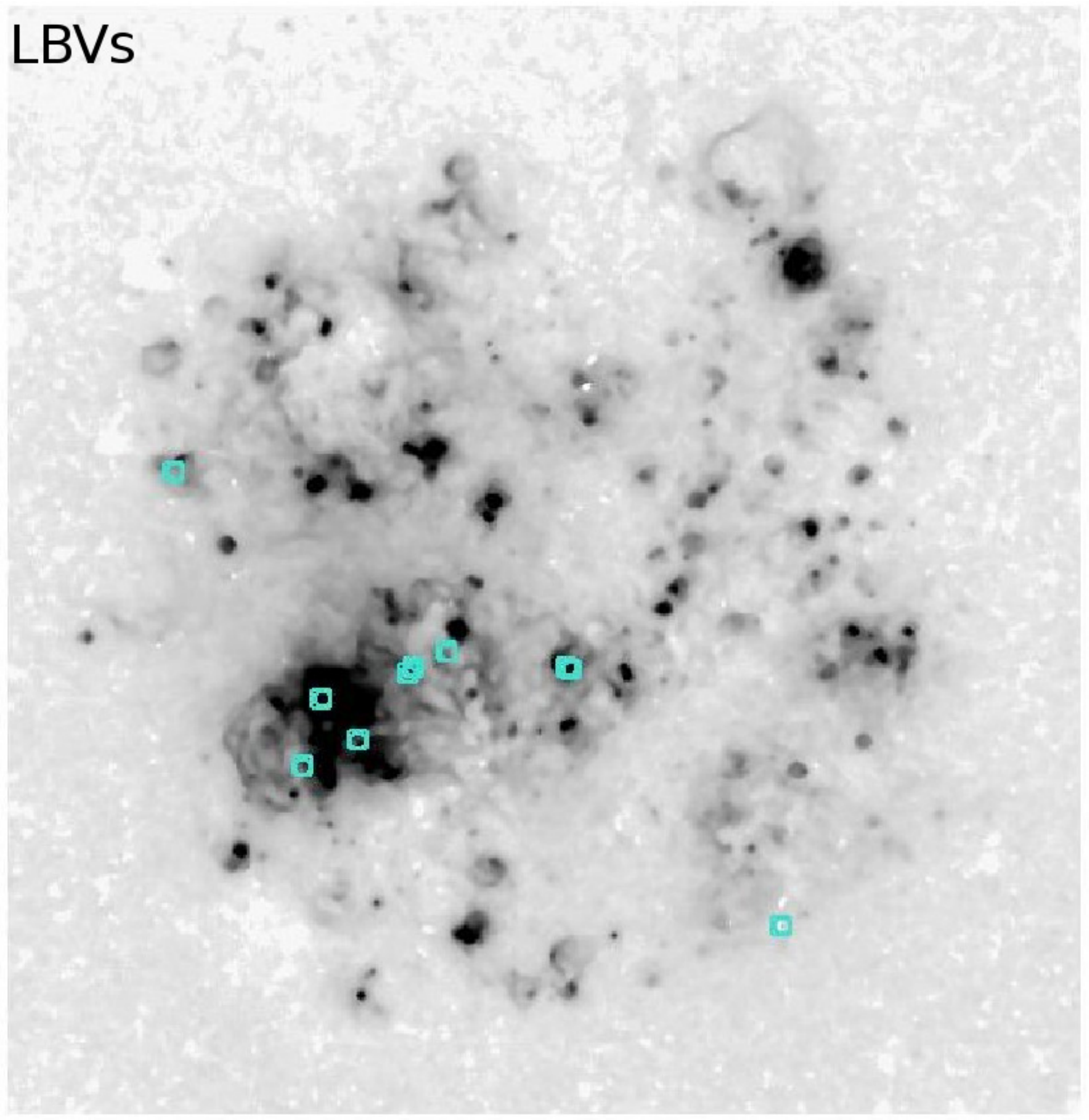}
\caption{H\,$\alpha$ images of the LMC, with our main-sequence B and O star samples as well as the SG B[e] stars and LBVs overplotted. A concentration of B stars is visible in a rectangular area (marked in red) with no strong H\,$\alpha$ sources, indicating a spatial bias in this subsample (see also Sect. 6). The black scale bar corresponds to 2 deg ($\sim$1.7 kpc).}
\label{fig:samples2}
\end{figure}

The coordinates and parameters (such as temperature, spectral type and/or luminosity) of the stellar samples in the two galaxies were obtained from various sources, described below. Because of the nature of our analysis (Sect. 3), the aim in each case was to obtain a list of coordinates with good \emph{spatial} coverage and to avoid biases against regions of high or low H\,$\alpha$ surface brightness. The individual catalogue papers contain details on the observations, reductions and the selection and confirmation of candidates in each catalogue; these will be briefly summarized below. The spatial coverage of the supergiant and WR catalogues in both galaxies is illustrated in Fig. \ref{fig:samples}, while the coverage of the main-sequence B and O stars, as well as SG B[e] stars and LBVs in the LMC, is shown in Fig. \ref{fig:samples2}.

\subsubsection{Supergiants}

A catalogue of spectroscopically classified RSGs and YSGs in M33 was compiled by Drout, Massey \& Meynet (2012). The observations were performed using the Hectospec multi-fibre spectrograph on the 6.5-m MMT telescope at the Fred Lawrence Whipple Observatory in Arizona, US. The RSG and YSG candidates in M33 were photometrically selected by Drout et al. from the Local Group Galaxy Survey \citep[LGGS;][]{lggs}, applying the following criteria: the YSG candidates had $V < 18.7$ mag, $U - B > -0.4$ mag and $0$ mag $\leq B - V \leq 1.4$ mag, while for the RSG candidates $V < 20.0$ mag and $V - R > 0.6$ mag. An additional criterion of $B -V > -1.599(V - R)^{2} + 4.18(V - R) - 0.83$ mag \citep{massey09} was adopted to separate likely foreground dwarfs from the supergiants. The rest of the foreground stars were excluded spectroscopically using the radial velocities of each star. Using the upper and lower temperature limits of YSGs set by \citet{m33sg} at 7500 K (the line between the A and F spectral classes) and 4800 K (the line between the G and K spectral classes), respectively, the catalogue contains 188 RSGs, 74 YSGs and 47 BSGs. Spatially, the catalogue covers the entire disk of M33, but only two RSGs in the central square kpc region are included because of crowding effects. As this region is quite small and does not in any way dominate the H\,$\alpha$ emission of M33 (Fig. \ref{fig:ha_pics}), we consider any biases in the spatial coverage with respect to H\,$\alpha$ surface brightness insignificant (this assumption is tested and found valid in Sect. 5). Luminosity-wise, the catalogue was argued by its authors to be complete down to a luminosity of log $L/L_{\odot} \sim 4.8$, and contains 154 stars below this limit as well.

A similar catalogue for the LMC was compiled by \citet{lmcsg} by observing 64 fields within the galaxy with the 138-fibre multi-object spectrometer Hydra on the Cerro Tololo 4-m telescope. YSG and RSG candidates were selected by Neugent et al. from the USNO CCD Astrograph Catalogue Part 3 (UCAC3) using magnitudes from the Two Micron All Sky Survey\footnote{http://www.ipac.caltech.edu/2mass/index.html} \citep[2MASS;][]{2mass} and spectroscopically classified. For RSG candidates $(J - K)_{\mathrm{2MASS}} < 0.9$ mag and $K_{\mathrm{2MASS}} \leq 10.2$ mag. For YSG candidates the $K$ brightness limit depends on $(J - K)$ as described in \citet{smcysg} -- at $(J - K)_{\mathrm{2MASS}} = 0$ mag the limit is $K_{\mathrm{2MASS}} = 12.6$ mag and at $(J - K)_{\mathrm{2MASS}} = 0.9$ mag it is $K_{\mathrm{2MASS}} = 10.2$ mag. Foreground red and yellow dwarf stars were excluded using radial velocities. Eight stars were classified merely as `possible LMC supergiants' (as opposed to `probable') and are excluded from the present study. Using the same limiting temperatures as for M33, the catalogue contains 543 RSGs, 109 YSGs and 163 BSGs. The spatial coverage of this catalogue, as demonstrated in \citet{lmcsg}, is not perfect over the entire galaxy but the 64 fields are spread roughly evenly over the galaxy and thus the catalogue should not be biased toward regions of any particular H\,$\alpha$ surface brightness. Such coverage is sufficient for this study (this assumption is also tested in Sect. 5). The catalogue was deemed by its authors to be missing no more than a few YSGs, while the RSG completeness is estimated to be roughly 30 per cent (out of some $\sim1800$ RSGs).

As the BSGs in the catalogues do not include stars with temperatures higher than $\sim11000$ K, we do not include the BSGs in our analysis.

\subsubsection{WR stars}

We have also obtained catalogues of spectroscopically classified WR stars in both galaxies. WR stars with CNO cycle products, especially nitrogen, on their surfaces, are classified WN, while those with triple-$\alpha$ products, especially carbon, on the surface are classified WC. WC stars are considered a later stage of evolution than WN, and stars that reach the WC stage before their deaths have a higher minimum initial mass \citep[$\sim$25 and $\sim$40 $M_{\odot}$ for WN and WC stars, respectively;][]{crowther07}. There is, however, also evidence \citep{georgy12,mcclelland16} that some WC stars require a different evolutionary channel than WN stars and may have initial masses lower than what \citet{crowther07} concluded. WN stars with strong hydrogen signatures are classified WNh \citep{smith96}. WNh stars are still considered to be core-hydrogen-burning, and their initial masses are believed to be higher than those of WN or WC stars \citep[$\gtrsim$60 $M_{\odot}$;][]{crowther07,smith08}.

For the LMC, we use the catalogue of WN stars by \citet{lmcwr}, considered by the catalogue authors to be complete (or very close to complete) over the entire galaxy. This catalogue is mostly based on the earlier BAT99 catalogue (Breysacher, Azzopardi \& Testor 1999), with some more recent re-classifications. This catalogue contains a total of 94 WN stars. For the WC stars, we use the massive star catalogue of \citet{bonanos}; the 24 WC stars, 22 of type WC4 and 2 labeled early-type WC, in this catalogue are also largely based on BAT99 and thought to represent a complete sample. For M33, we use the catalogue of WR stars by \citet{m33wr}. This catalogue is considered by its authors to be $\sim 95$ per cent complete, and contains 139 WN stars, 52 WC stars and 10 Ofpe/WN9 stars. Stars with an uncertain WR status (classification for example `WN?') are excluded -- there is one of these in the LMC and three in M33.

\subsubsection{Other evolved stars}

We also use the LBV stars listed by \citet{smith15}. These include all confirmed LBVs in the LMC and likely, well-studied LBV candidates with shells. We exclude the five unrelated stars and the duplicate entry pointed out by \citet{antismith}, bringing the total number to ten LBVs -- three classical LBVs and seven lower-luminosity LBVs. In M33, there are only four confirmed LBVs \citep[Hubble-Sandage variables B, C and 2 and Var 83;][]{lbvm33}, but we include them as well \citep[the actual number of LBV stars in M33 was estimated to be in the hundreds by][]{massey07}. As \citet{smith15} considered supergiant B[e] (SG B[e]) stars to be the lower-mass analogs of LBVs, these stars in the LMC are included for comparison. The coordinates of 12 spectroscopically classified SG B[e] stars were obtained from \citet{zickgraf} and \citet{bonanos}; this includes every such star discovered in the LMC.

\subsubsection{Main-sequence stars}

As well as these evolved stars, we analyze massive main-sequence stars in the LMC with spectroscopic classification. Their coordinates were obtained from the SIMBAD Astronomical Database\footnote{http://simbad.u-strasbg.fr/simbad/} \citep{simbad}, with a search radius of 5 degrees from the centre of the LMC as reported in NED. With such a search we have obtained the coordinates of main-sequence stars between the B2V and O3V spectral types as reported in April 2015. Very few stars classified as main-sequence spectral types later than B2V can be found in the LMC (for example, only 5 stars of type B3V as opposed to 92 of type B2V). The main-sequence subsamples are most likely spatially biased, and are only used for qualitative purposes in this paper. The \citet{bonanos} catalogue contains hundreds of main-sequence stars, but being a compilation of catalogues that mostly target specific regions in the LMC, as a main-sequence catalogue it is also spatially biased. As the SIMBAD search results simply contain higher numbers of stars, we use them instead.

\subsection{SN samples}

To compare the pixel statistics (Sect. 3) of the stars to previous statistics of CCSNe, we use the distances and results of the sample of SNe reported in A12 which include the results of AJ08 as well. The host galaxies of the SNe in AJ08 and A12 were observed using the 1-m Jacobus Kapteyn Telescope (JKT), the 2-m Liverpool Telescope (LT) and the 2.5-m Isaac Newton Telescope (INT) at the Observatorio del Roque de los Muchachos on La Palma, Canary Islands, and with the European Southern Observatory (ESO) 2.2-m telescope at La Silla, Chile. In some cases we also make use of the results of K13. The H\,$\alpha$ images in K13 were observed with the Andalucia Faint Object Spectrograph and Camera (ALFOSC) on the 2.5-m Nordic Optical Telescope (NOT) on La Palma. Details of these observations can be found in the aforementioned papers.

Samples of SNe are potentially affected by issues in precise SN classification. Type Ib, Ic and IIb SNe can be difficult to distinguish spectroscopically \citep{sntypes}. Although type II-L SNe generally exhibit a weaker P Cygni absorption profile at the H$\alpha$ line than type II-P SNe \citep{schlegel96}, separating these types can be difficult without light curves which are not always available -- not to mention that e.g. \citet{a14} argue that there is no unambiguous borderline between types II-P and II-L at all. The A12 SN subsamples may therefore be affected by mixing between the subtypes (for example, one type Ib/IIb SN was included as half a type Ib and half a type IIb). Type IIn SNe, furthermore, are a very diverse class with several possible progenitor channels, with narrow hydrogen emission lines being the common feature. Therefore the sample of type IIn SNe may not include representative fractions of different SNe within this type. These effects may result in, for example, larger scatter in the correlations between SNe and H\,$\alpha$ emission.

%

\section{Analysis}

Here we describe the methods we use to connect CCSNe to possible progenitor populations.

\subsection{The pixel statistics method}

The pixel statistics analysis method was introduced by \citet{fruchter06}, who applied it to blue light to study the long gamma-ray burst distribution, and JA06. It is described in detail by JA06, and has since been used to study correlations between CCSNe and emission at different wavelengths, mainly H\,$\alpha$, by AJ08, A12, K13 and H14.

Briefly put, each pixel in the H\,$\alpha$ image of a galaxy is assigned a number, called its normalized cumulative rank (NCR), that tells us what fraction of the H\,$\alpha$ emission originates in pixels with lower surface brightness. When the coordinates of a SN or, in the case of this study, a star, are connected to a pixel and thus an NCR, this acts as an indicator of any correlation between the SN or star and the underlying H\,$\alpha$ emission. Each `background' pixel with no emission has an NCR of 0, while the brightest pixel of the galaxy has an NCR of 1. This requires that the bias level and sky emission have been properly subtracted. When NCRs are calculated for a sample of objects, the mean NCR and distribution of the NCRs can be used to statistically compare the samples, for example, with an Anderson-Darling (AD) test. A distribution that exactly follows the underlying emission has a mean value of 0.5 and a uniform distribution. As an example, 70 per cent of the sample will have NCR $\leq 0.7$ since 70 per cent of the emission comes from pixels with NCR $\leq 0.7$. 

In earlier studies, starting with JA06, a higher NCR(H\,$\alpha$) for a set of SNe has been interpreted as a shorter lifespan of the SN progenitor and thus an initially more massive progenitor star. The studies of e.g. A12 and K13 probed structures at a projected scale of roughly 100 or 200 pc depending on the distance of the galaxy. Any correlation with H\,$\alpha$ emission thus does not indicate a correlation with small H~{\sc ii} regions created by individual massive stars but with larger H~{\sc ii} complexes. \citet{crowther13} argued that the ages of these complexes are measured in tens of Myr, as opposed to smaller H~{\sc ii} regions which only last as long as their central star continues to emit ionizing radiation. The lifespan of a large H~{\sc ii} complex is thus comparable to that of a star with initial mass around 10 $M_{\odot}$ and the same complex can host more than one generation of more massive stars. 

Nonetheless, as a star grows older, a few things happen to lower its NCR. The star gradually drifts away from its native H~{\sc ii} complex, and the complex may grow fainter or dissipate entirely as massive stars within it explode as SNe, sweeping the gas away, and stop emitting ionizing radiation. Therefore even though a relatively low-mass SN progenitor may still be inside its H~{\sc ii} complex when it explodes, it is still \emph{less likely} to be there. Some such stars have a high NCR, being born early in the life of the complex or having drifted into another H~{\sc ii} region. Others, born at later times, have a lower (possibly zero) NCR as the complex disappears around them -- the timing depends on the lifespan of the star. On the other hand, the higher-mass stars have the shortest lifetimes and the highest fluxes of ionizing radiation, and stay closer to the centre of the H~{\sc ii} region. 

Some stars are born into smaller H~{\sc ii} regions than others. Because complexes of different size and brightness have different NCRs, and some smaller and fainter regions are obscured by noise in the H\,$\alpha$ image, even high-mass stars occasionally have low NCRs. The smallest H~{\sc ii} regions also have the shortest lifespans, which may further lower the NCRs of stars born in them although, on the spatial scales probed by A12 and K13, the small H~{\sc ii} regions either blend into the large ones or do not have a high NCR to begin with. Conversely, projection effects may result in higher NCRs for stars lying on the line of sight to an H~{\sc ii} complex. However, statistically, the longer the star lives the less likely it is to have a high NCR. Therefore, samples of stars or SNe should still form a sequence of descending initial mass with descending NCR. Testing this assumption is one of the aims of this paper.

Some [N~{\sc ii}] $\lambda\lambda6548,6584$ emission is included in the H\,$\alpha$ filters used in the SHASSA survey \citep{lmcpic} and by \citet{m33pic}. The level of contamination differs with the metallicity of the emitting region, and metallicity gradients in galaxies could bias NCR distributions. However, using Eq. 6 in \citet{helm04}, we estimate that the [N~{\sc ii}]/H\,$\alpha$ flux ratios in the LMC and M33 should be about 0.14 and 0.2, respectively. Furthermore, because of the transmission curve of the filter used by \citet{lmcpic}, only a third of the [N~{\sc ii}] emission would be observed, while the peak transmission (at the H\,$\alpha$ line) was 78 per cent. For M33, \citet{m33pic} estimated the [N~{\sc ii}] contribution in their study to be 5 per cent or less of the total flux. Therefore, metallicity differences between the two galaxies or between regions inside one galaxy should not significantly affect observed fluxes in the H\,$\alpha$ filters or the resulting NCRs. Previous NCR studies of SNe, which we use for comparison, were similarly contaminated. The metallicity difference between galaxies is not a concern because NCRs are by definition normalized to the brightest region in a galaxy. Metallicity-related NCR biases should be mainly due to internal metallicity gradients (or their absence) inside galaxies. Assuming that the [N~{\sc ii}] lines mainly fall on the wings of the H\,$\alpha$ filter, we conclude that the effect on our NCR comparisons should be minimal. Line strengths in H~{\sc ii} regions are also affected by other factors, such as differences in dust content, ionization, temperature and density in the regions, so that H\,$\alpha$ emission does not perfectly trace star formation. However, estimating the biases these effects may cause in the NCRs is outside the scope of this paper. For now, we neglect them for simplicity, consistently with other NCR studies.

\begin{figure}
\centering
\includegraphics[width=\columnwidth]{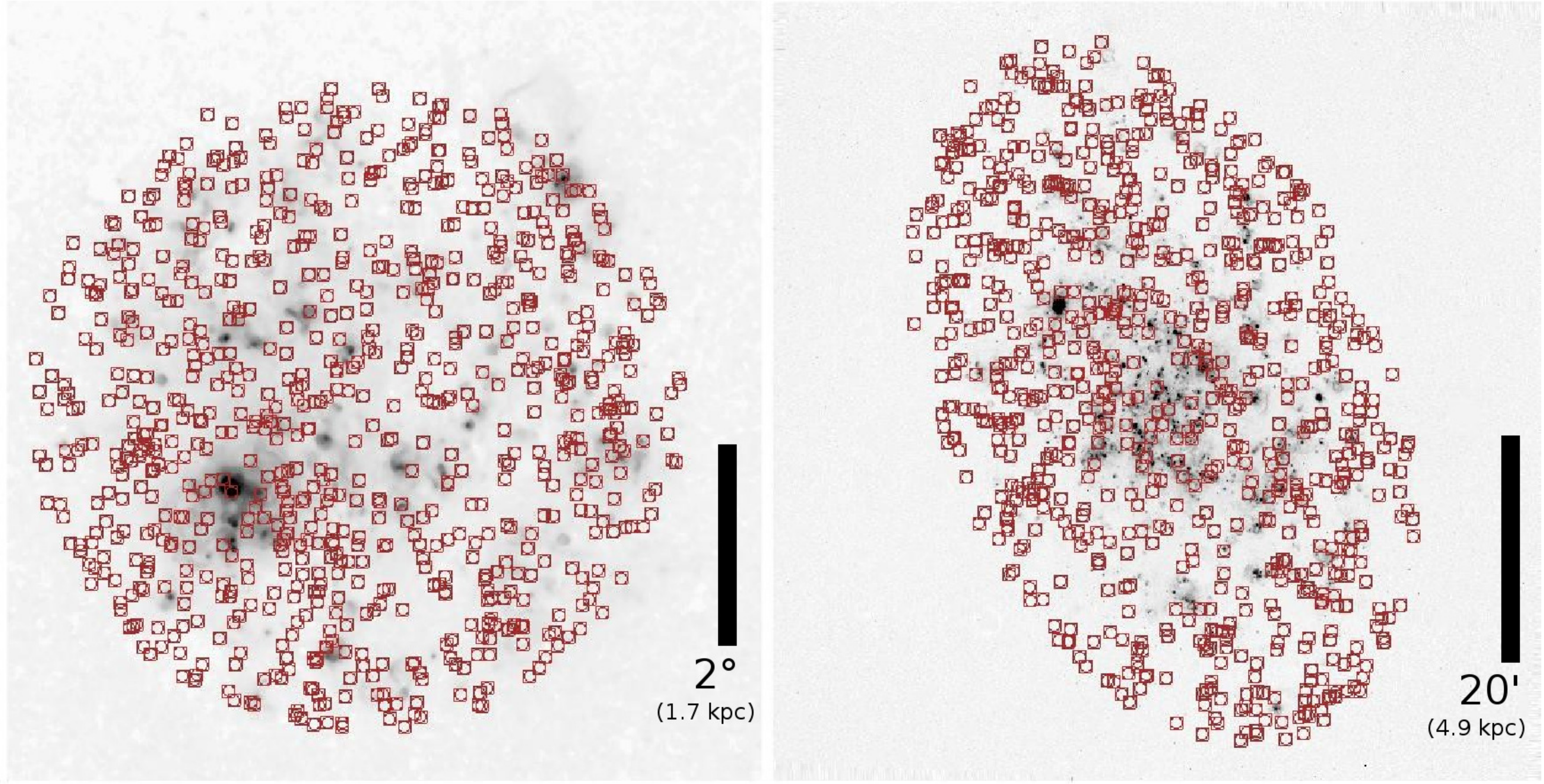}
\caption{H\,$\alpha$ images of the LMC (left) and M33 (right), with the coordinates of our randomly generated uniformly distributed samples of positions overplotted.}
\label{fig:randoms}
\end{figure}

With this in mind, we calculate the NCRs of each star in each of our samples. For comparison, we also generate a sample of 250 random positions within each of the two galaxies, with a uniform spatial distribution inside the visible extent of the galaxy. For the LMC this corresponds to a circle with a radius of 3.5 deg, and for M33 an ellipse with $a = 33$ arcmin, $b = 20$ arcmin and position angle 22.7 deg. The distribution of the random positions is shown in Fig.~\ref{fig:randoms}. These steps are repeated for each of the images we analyze in this study, as described below.

\subsection{Comparison to CCSN pixel statistics}

In order to make our NCR results comparable to those of the SNe in A12, we simulate observations of the two galaxies at a distance, spatial resolution and signal-to-noise level comparable to the galaxies in A12. All the galaxy images in this sample were re-binned to a pixel scale of $\sim$1 arcsec pix$^{-1}$. The median distances of the type II-P and II-L SN host galaxies in A12 are close to 20 Mpc and those of type Ib, Ic, IIn and IIb SNe are all close to 35 Mpc. Therefore we simulate distances of 20 and 35 Mpc, and using these we obtain our main results. The images are first convolved to have a spatial resolution (FWHM) close to what 1 arcsec would correspond to at distances of 20 and 35 Mpc, that is 97 and 170 pc, respectively. This resolution is comparable to the sizes of giant H~{\sc ii} regions \citep{crowther13}. These are equivalent to 8.4 and 14.7 pix in the original LMC image and 11.8 and 20.6 pix in the M33 image. The images are then re-binned to have a pixel scale as close as possible to 1 arcsec pix$^{-1}$ at this simulated distance. This is done using the {\sc iraf} tasks {\sc gauss} and {\sc blkavg}. In addition, using the task {\sc mknoise} in the package {\sc artdata}, Poisson noise is added to the images in order to match the signal-to-noise ratio (SNR) per pixel with the images used by A12. The SNR per pixel of the images used by A12 was on the order of 50 to 100 in H~{\sc ii} regions of average brightness and on the order of 10 to 20 in faint ones.

To make the spatial scales of the H\,$\alpha$ images comparable to the images in A12, the LMC image is re-binned by 8 to simulate a distance of 20 Mpc and by 15 for 35 Mpc, while the M33 image is re-binned by 12 and by 21 for these distances. In the original image, this re-binning would correspond to a pixel scale of over 6 arcmin pix$^{-1}$ for the LMC and over 24 arcsec pix$^{-1}$ for M33. The possible uncertainties in the coordinates of the stars in these galaxies, or in the coordinate systems of the images themselves (on the order of arcseconds or less depending on the galaxy), are insignificant compared to this pixel size. Therefore the positions and NCRs of the stars in our samples are sufficiently accurate. However, the uncertainties in SN positions and in the coordinate systems of the images of their host galaxies are up to $\sim$1 arcsec (as was conservatively estimated by K13), which is comparable to the pixel size of the images of A12. Therefore the NCRs of the SNe and local galaxy stars should not be directly compared. To compensate for this, we simulate how an accurate NCR distribution would change with the addition of positional uncertainties by performing the following Monte Carlo analysis. After calculating the NCRs of each stellar sample using the original, accurate coordinates, an offset with a random direction and distance is added to each coordinate set. The offset distances are drawn from a Gaussian probability distribution with a standard deviation of $\sigma = 0.5$ arcsec. This accounts for the errors in the world coordinate system fitting typical for the images in K13 (0.2 arcsec) and uncertainties in the coordinates reported for SNe. Thus 95 per cent of the positional shifts are $2\sigma = 1$ arcsec or less. This is repeated 1000 times for each stellar sample and a new mean NCR is then calculated. The errors of the NCRs with accurate coordinates are reported as standard error of the mean (SEM), while the errors reported for the NCRs with positional errors are $\frac{\sigma_{\mathrm{MC}}}{\sqrt{N}}$, where $\sigma_{\mathrm{MC}}$ is the standard deviation of the Monte Carlo NCRs for the sample and $N$ the original size of the stellar sample. This is analogous to the SEM in order to be comparable to the SEMs reported by A12 and K13. The NCR distributions with offsets can then finally be compared to those of the SNe. Both the accurate NCRs and those calculated with offsets are listed in Sect. 4.

The distributions of NCRs that include positional errors should not be compared to others with an AD test using the Monte Carlo distribution as it is -- despite the Monte Carlo distribution having 1000 times more individual values, the analysis does not increase information. Therefore, before any AD tests with these distributions are made, we re-bin them by sorting them in order of increasing NCR and taking the average of every 1000 values.

\begin{table*}
\centering
\begin{minipage}{140mm}
\caption{Mean NCRs ($<$NCR$>$) of different stellar types in the LMC at a simulated distance of 20 and 35 Mpc, with accurate coordinates (`acc') and simulated positional errors of $\sigma = 0.5$ arcsec (`err'). The indicative initial masses of main-sequence stars are from \citet{wv10} for O stars and from \citet{allen} for B stars.}
\label{table:main_ncr_lmc}
\begin{tabular}[t]{lccccc}
    \hline
        & & \multicolumn{2}{c}{20 Mpc} & \multicolumn{2}{c}{35 Mpc} \\
	Stellar type & N & $<$NCR$>$(acc) & $<$NCR$>$(err) & $<$NCR$>$(acc) & $<$NCR$>$(err) \\
    \hline
	random & 250 & 0.095 $\pm$ 0.007 & - & 0.101 $\pm$ 0.006 & - \\
	B2V (8 $M_{\odot}$) & 92 & 0.486 $\pm$ 0.030 & 0.472 $\pm$ 0.030 & 0.524 $\pm$ 0.030 & 0.513 $\pm$ 0.029 \\	
	B1V (13 $M_{\odot}$) & 135 & 0.530 $\pm$ 0.024 & 0.509 $\pm$ 0.025 & 0.537 $\pm$ 0.023 & 0.529 $\pm$ 0.023 \\		
	B0V (17.5 $M_{\odot}$) & 147 & 0.627 $\pm$ 0.022 & 0.610 $\pm$ 0.023 & 0.631 $\pm$ 0.022 & 0.603 $\pm$ 0.022 \\
	O9V (20 $M_{\odot}$) & 117 & 0.692 $\pm$ 0.022 & 0.658 $\pm$ 0.023 & 0.646 $\pm$ 0.023 & 0.615 $\pm$ 0.023 \\	
	O8V (25 $M_{\odot}$)& 89 & 0.667 $\pm$ 0.028 & 0.637 $\pm$ 0.029 & 0.623 $\pm$ 0.030 & 0.594 $\pm$ 0.030 \\	
	O7V (31 $M_{\odot}$) & 52 & 0.719 $\pm$ 0.027 & 0.685 $\pm$ 0.029 & 0.678 $\pm$ 0.030 & 0.645 $\pm$ 0.031 \\	
	O6V (37 $M_{\odot}$) & 48 & 0.742 $\pm$ 0.031 & 0.711 $\pm$ 0.034 & 0.706 $\pm$ 0.035 & 0.673 $\pm$ 0.036 \\		
	O5V (44 $M_{\odot}$)& 13 & 0.805 $\pm$ 0.065 & 0.776 $\pm$ 0.060 & 0.785 $\pm$ 0.061 & 0.755 $\pm$ 0.063 \\		
	O4V (53 $M_{\odot}$) & 14 & 0.820 $\pm$ 0.073 & 0.784 $\pm$ 0.075 & 0.792 $\pm$ 0.085 & 0.745 $\pm$ 0.083 \\	
	O3V (64 $M_{\odot}$) & 12 & 0.961 $\pm$ 0.027 & 0.931 $\pm$ 0.030 & 0.952 $\pm$ 0.034 & 0.911 $\pm$ 0.037 \\
	RSG & 543 & 0.182 $\pm$ 0.010 & 0.180 $\pm$ 0.010 & 0.229 $\pm$ 0.010 & 0.228 $\pm$ 0.010 \\
	~RSG (log $L/L_{\odot} < 4.6$) & 361 & 0.155 $\pm$ 0.011 & 0.152 $\pm$ 0.011 & 0.196 $\pm$ 0.011 & 0.196 $\pm$ 0.011 \\
	~RSG (log $L/L_{\odot} \ge 4.6$) & 182 & 0.236 $\pm$ 0.018 & 0.239 $\pm$ 0.018 & 0.295 $\pm$ 0.017 & 0.290 $\pm$ 0.017 \\
 	~RSG (log $L/L_{\odot} \ge 4.8$) & 76 & 0.267 $\pm$ 0.031 & 0.268 $\pm$ 0.031 & 0.321 $\pm$ 0.029 & 0.321 $\pm$ 0.029 \\
	YSG & 109 & 0.331 $\pm$ 0.029 & 0.328 $\pm$ 0.029 & 0.387 $\pm$ 0.028 & 0.375 $\pm$ 0.028 \\
	~YSG (log $L/L_{\odot} \ge 4.8$) & 37 & 0.373 $\pm$ 0.044 & 0.362 $\pm$ 0.044 & 0.417 $\pm$ 0.047 & 0.412 $\pm$ 0.044 \\	
	SG B[e] & 12 & 0.340 $\pm$ 0.086 & 0.342 $\pm$ 0.079 & 0.371 $\pm$ 0.083 & 0.375 $\pm$ 0.082 \\
	LBV & 10 & 0.523 $\pm$ 0.082 & 0.511 $\pm$ 0.075 & 0.539 $\pm$ 0.085 & 0.527 $\pm$ 0.080 \\
	 ~Classical LBV & 3 & 0.774 $\pm$ 0.115 & 0.750 $\pm$ 0.086 & 0.785 $\pm$ 0.110 & 0.761 $\pm$ 0.096 \\	
	 ~Low-luminosity LBV & 7 & 0.416 $\pm$ 0.077 & 0.409 $\pm$ 0.072 & 0.434 $\pm$ 0.087 & 0.427 $\pm$ 0.081 \\	
	WN & 94 & 0.561 $\pm$ 0.031 & 0.544 $\pm$ 0.032 & 0.575 $\pm$ 0.032 & 0.553 $\pm$ 0.032 \\
	 ~Early WN & 67 & 0.508 $\pm$ 0.035 & 0.490 $\pm$ 0.036 & 0.525 $\pm$ 0.036 & 0.503 $\pm$ 0.036 \\
	 ~Late WN & 27 & 0.676 $\pm$ 0.058 & 0.663 $\pm$ 0.057 & 0.684 $\pm$ 0.059 & 0.665 $\pm$ 0.057 \\
	WN (no H)& 45 & 0.515 $\pm$ 0.043 & 0.492 $\pm$ 0.044 & 0.517 $\pm$ 0.043 & 0.502 $\pm$ 0.043 \\
	 ~Early WN (no H) & 38 & 0.442 $\pm$ 0.039 & 0.419 $\pm$ 0.041 & 0.442 $\pm$ 0.039 & 0.430 $\pm$ 0.040 \\
	 ~Late WN (no H) & 7 & 0.847 $\pm$ 0.081 & 0.832 $\pm$ 0.073 & 0.866 $\pm$ 0.070 & 0.821 $\pm$ 0.076 \\
	(Early) WC & 24 & 0.656 $\pm$ 0.045 & 0.641 $\pm$ 0.045 & 0.662 $\pm$ 0.048 & 0.632 $\pm$ 0.050 \\
\hline
\end{tabular}
\end{minipage}
\end{table*}

\begin{table*}
\centering
\begin{minipage}{140mm}
\caption{As Table \ref{table:main_ncr_lmc}, but for M33.}
\label{table:main_ncr_m33}
\begin{tabular}[t]{lccccc}
    \hline
        & & \multicolumn{2}{c}{20 Mpc} & \multicolumn{2}{c}{35 Mpc} \\
	Stellar type & N & $<$NCR$>$(acc) & $<$NCR$>$(err) & $<$NCR$>$(acc) & $<$NCR$>$(err) \\
    \hline
	random & 250 & 0.089 $\pm$ 0.007 & - & 0.111 $\pm$ 0.008 & - \\	
	RSG & 188 & 0.203 $\pm$ 0.018 & 0.201 $\pm$ 0.018 & 0.237 $\pm$ 0.018 & 0.228 $\pm$ 0.017 \\
	 ~RSG (log $L/L_{\odot} < 4.6$) & 68 & 0.095 $\pm$ 0.022 & 0.097 $\pm$ 0.021 & 0.137 $\pm$ 0.021 & 0.137 $\pm$ 0.023 \\
	 ~RSG (log $L/L_{\odot} \ge 4.6$) & 120 & 0.261 $\pm$ 0.024 & 0.260 $\pm$ 0.023 & 0.293 $\pm$ 0.023 & 0.243 $\pm$ 0.023 \\
 	 ~RSG (log $L/L_{\odot} \ge 4.8$) & 69 & 0.330 $\pm$ 0.033 & 0.324 $\pm$ 0.032 & 0.361 $\pm$ 0.032 & 0.339 $\pm$ 0.031 \\
	YSG & 74 & 0.358 $\pm$ 0.029 & 0.353 $\pm$ 0.030 & 0.412 $\pm$ 0.030 & 0.399 $\pm$ 0.030 \\
	 ~YSG (log $L/L_{\odot} \ge 4.8$) & 58 & 0.389 $\pm$ 0.032 & 0.379 $\pm$ 0.033 & 0.440 $\pm$ 0.033 & 0.423 $\pm$ 0.033 \\		
	LBV & 4 & 0.480 $\pm$ 0.080 & 0.496 $\pm$ 0.078 & 0.492 $\pm$ 0.060 & 0.508 $\pm$ 0.077 \\
	Ofpe/WN9 & 10 & 0.589 $\pm$ 0.096 & 0.573 $\pm$ 0.091 & 0.532 $\pm$ 0.089 & 0.533 $\pm$ 0.089 \\
	WN & 139 & 0.592 $\pm$ 0.022 & 0.566 $\pm$ 0.022 & 0.576 $\pm$ 0.022 & 0.551 $\pm$ 0.023 \\
	 ~Early WN & 81 & 0.516 $\pm$ 0.027 & 0.491 $\pm$ 0.027 & 0.490 $\pm$ 0.026 & 0.464 $\pm$ 0.027 \\
	 ~Late WN & 38 & 0.740 $\pm$ 0.041 & 0.711 $\pm$ 0.041 & 0.734 $\pm$ 0.043 & 0.704 $\pm$ 0.043 \\
	WC & 52 & 0.572 $\pm$ 0.032 & 0.552 $\pm$ 0.033 & 0.585 $\pm$ 0.034 & 0.558 $\pm$ 0.034 \\
	 ~Early WC & 28 & 0.551 $\pm$ 0.049 & 0.522 $\pm$ 0.048 & 0.536 $\pm$ 0.050 & 0.516 $\pm$ 0.050 \\
	 ~Late WC & 14 & 0.665 $\pm$ 0.042 & 0.663 $\pm$ 0.043 & 0.728 $\pm$ 0.039 & 0.692 $\pm$ 0.042 \\
	
\hline
\end{tabular}
\end{minipage}
\end{table*}

\section{Results}

We divide our samples from catalogues and SIMBAD (described in Sect. 2) into sub-samples as follows. Each of the spectral types of the LMC main-sequence stars is its own sub-sample. The Wolf-Rayet samples are split into WN and WC stars in both galaxies. The WN stars are split into `early'- (WN2-WN5) and `late'-type (WN6+) stars. In the LMC we analyze samples of WN stars including WNh stars, simply labeled "WN", and without WNh stars, labeled "WN (no H)". While WN and WNh stars are expected to be systematically different \citep{crowther07, smith08}, this approach eases the comparison between the two galaxies. In M33, the \citet{m33wr} catalogue only contains one WNh star. However, \citet{hamann06} listed many WNh stars in the Milky Way, providing circumstantial evidence that they should also exist in larger numbers in M33; one star in the \citet{m33wr} catalogue is labeled "H-rich" but not classified WNh; one of the classification sources \citep{abbott04} intentionally omitted the \citet{smith96} WNh criteria; and one WN7h star from another source \citep{drissen08} is simply reported as a WN7 in \citet{m33wr}. Therefore, in M33, WN and WNh stars seem to be mixed and cannot be reliably separated. WC stars are also split into early- and late-type; however, in the LMC, the WC stars are all early-type. In M33, we also analyze the ten stars listed as `Ofpe/WN9' as their own sub-sample. We set a cutoff in the RSG sample at log $L/L_{\odot} \ge 4.6$, close to the luminosity of RSGs with initial mass $9 M_{\odot}$ \citep{smartt+09}, which is consistent with the lower limits of type II-P progenitors estimated by \citet{smartt+09} and \citet{smartt15}. Another cutoff is set at log $L/L_{\odot} \ge 4.8$, which is the completeness limit of the M33 supergiant catalogue by \citet{m33sg}, for the purpose of comparison between the two galaxies. We also split the YSG samples at log $L/L_{\odot} = 4.8$ for the same purpose.

The mean NCRs of the different stellar samples, with both accurate coordinates and simulated positional errors, are presented in Table \ref{table:main_ncr_lmc} for the LMC and Table \ref{table:main_ncr_m33} for M33. The errors reported for the mean NCRs are the SEM in the case of accurate coordinates and, as explained in Sect. 3, the analogous $\frac{\sigma_{\mathrm{MC}}}{\sqrt{N}}$ from the Monte Carlo analysis. The mean NCR values for different types of transient events from A12 and H14 are presented in Table~\ref{table:sn_ncr}. 

\begin{figure}
\centering
\includegraphics[width=\columnwidth]{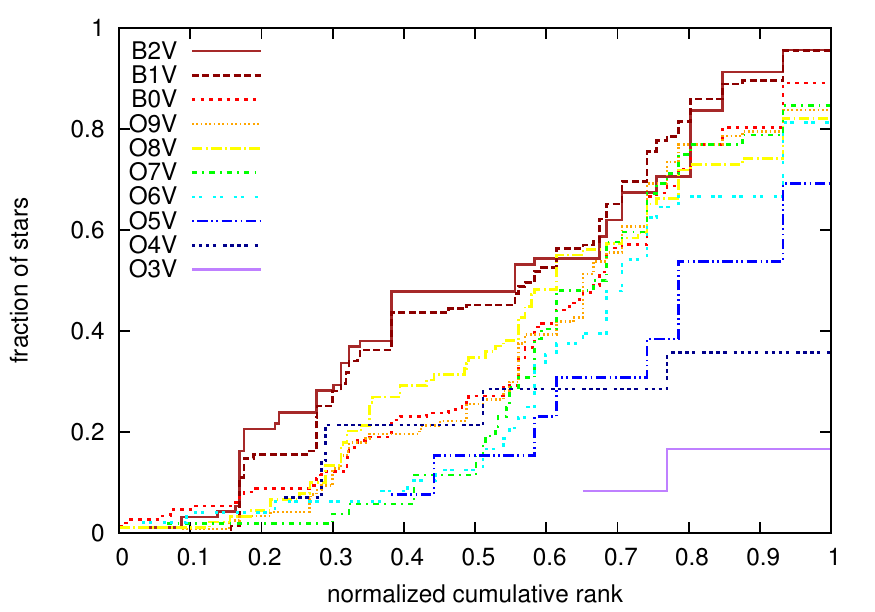}
\includegraphics[width=\columnwidth]{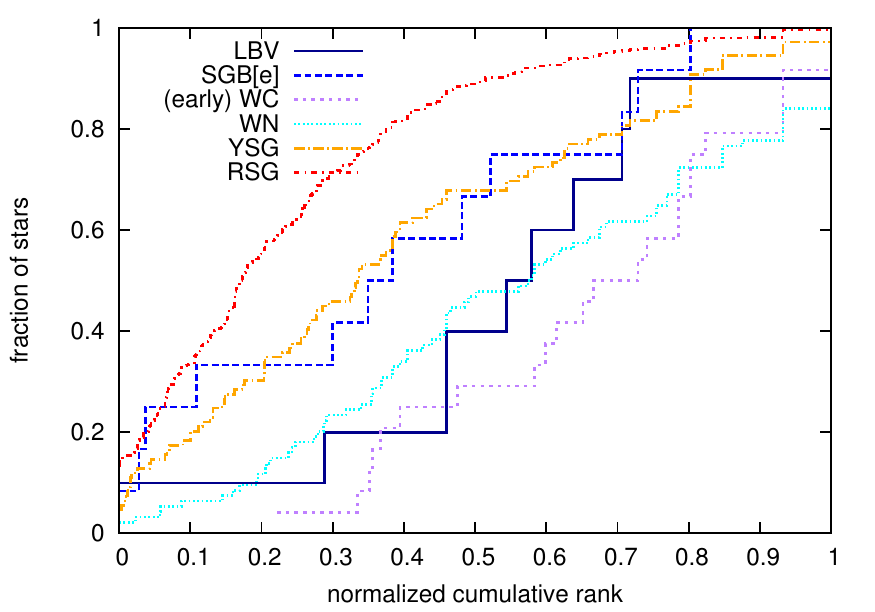}
\caption{Top panel: the cumulative NCR distributions of the main sequence stars of different spectral type in the LMC at a simulated distance of 35 Mpc, illustrating the trend of increasing NCR with earlier spectral type. Bottom panel: the distributions of different types of evolved stars in the LMC at the distance of 35 Mpc, similarly illustrating the dependence of NCR on stellar type and hence initial mass. In both panels, shorter-lived stars tend toward the lower right and longer-lived stars toward the upper left.}
\label{fig:star_ncr}
\end{figure}

Within the main-sequence sample in the LMC, it is evident that, in most cases, stars with an earlier spectral type -- and thus higher initial mass -- have a higher mean NCR, apart from the O8V and O9V stars, whose NCRs are consistent within the uncertainties. The mass-NCR correspondence is also visible when comparing evolved stars. Wolf-Rayet stars, with a lower initial mass limit around 20 or 25 $M_{\odot}$ \citep{crowther07,lmcwr}, have a substantially higher mean NCR than the supergiants which mostly occupy a lower initial mass range. For example, at 20 Mpc in the LMC, the WN sample has a mean NCR of 0.561 $\pm$ 0.031 while the RSG sample has 0.182 $\pm$ 0.010. Setting a higher luminosity cutoff (corresponding to higher minimum initial mass) to the supergiant samples also results in a higher mean NCR. For example, at 20 Mpc in the M33, the RSGs with log $L/L_{\odot} \ge 4.6$ have a mean NCR of 0.261 $\pm$ 0.024 while those with log $L/L_{\odot} \ge 4.8$ have 0.330 $\pm$ 0.033. Fig.~\ref{fig:star_ncr} shows the cumulative distributions of the main-sequence and evolved stars (with accurate coordinates) in the LMC at a simulated distance of 35 Mpc to illustrate the mass-NCR sequence.

\subsection{Differences between galaxies}

\begin{table}
\centering
\caption{Mean NCRs ($<$NCR$>$) of different CCSN types and impostors from the literature, and median distances of each transient sample. The type Ic SNe in K13 are located in strongly star-forming galaxies. The results of other SN types in K13 are consistent with A12. The redshifts (radial velocities) of the host galaxies of the transients are obtained from these references if given, and from the NED if not, and converted to distances using $H_{0} = 70$ km s$^{-1}$ Mpc$^{-1}$, $\Omega_{m} = 0.3$, $\Omega_{\Lambda} = 0.7$.}
\label{table:sn_ncr}
\begin{tabular}[t]{lcccc}
    \hline
        Transient & N & $<$NCR$>$ & Median distance & Ref. \\
	 & & & (Mpc) & \\
    \hline
	II-P & 58 & 0.264 $\pm$ 0.039 & 21.4 & A12 \\
	II-L & 13 & 0.375 $\pm$ 0.102 & 17.8 & A12 \\
	IIb & 13.5 & 0.402 $\pm$ 0.095 & 35.1 & A12 \\
	Ib & 39.5 & 0.318 $\pm$ 0.045 & 40.5 & A12 \\
	Ic & 52 & 0.469 $\pm$ 0.040 & 35.3 & A12 \\
	Ic & 18 & 0.607 $\pm$ 0.068 & 26.3 & K13 \\
	IIn & 24 & 0.225 $\pm$ 0.058 & 39.3 & H14 \\
	Impostor & 13 & 0.133 $\pm$ 0.086 & 8.5 & H14 \\
\hline
\end{tabular}
\end{table}

Generally speaking, the NCR results in Tables \ref{table:main_ncr_lmc} and \ref{table:main_ncr_m33} are quite similar between the two galaxies used in this study, despite the differences between the galaxies. M33 is a late-type spiral galaxy with a metallicity gradient between roughly solar metallicity ($Z_{\odot}$) in the nucleus and $0.5Z_{\odot}$ in the outer disk \citep{m33metal}. Its baryonic mass is about $10^{10}$ $M_{\odot}$ \citep{m33mass}. The LMC is a dwarf spiral galaxy with a baryonic mass about 3 $\times$ $10^{9}$ $M_{\odot}$ \citep{lmcmass} and exhibits signs of a bar structure. The LMC has a metallicity generally around $0.4Z_{\odot}$ without a clear gradient \citep{lmcmetal}. 

The mean NCRs of the RSG samples with log $L/L_{\odot} \ge 4.6$ and log $L/L_{\odot} \ge 4.8$ are inconsistent within a 1 $\sigma$ limit but consistent within 2 $\sigma$. They are also formally consistent according to the AD test: the probabilities for having identical distributions are 41 and 29 per cent respectively using the 20 Mpc distance. Because of completeness issues, the RSGs with log $L/L_{\odot} \le 4.6$ in M33 are biased toward lower values than in the LMC (0.095 $\pm$ 0.022 and 0.155 $\pm$ 0.011, respectively, at 20 Mpc). This is probably caused by a larger fraction of missing low-luminosity RSGs against bright regions in M33. The YSGs with log $L/L_{\odot} \ge 4.8$ have both consistent mean NCRs and a 72 per cent AD test probability. The mean NCRs of WN-star samples (including WNh stars) are also self-consistent, with AD test probabilities of 23 and 26 per cent for early and late WN respectively. There is thus no apparent metallicity effect on the WN progenitors, in agreement with \citet{lmcwr}, who find a lower mass limit for WN stars in the LMC ($\sim$20 $M_{\odot}$) comparable to the limit in the Milky Way ($\sim$25 $M_{\odot}$). They suggest that this may be because the reduced mass loss caused by low metallicity implies less angular momentum loss and thus faster rotation, leading to lower minimum WR star mass.

The only significant difference between the galaxies concerns the early WC stars, which have a lower NCR in M33 than in the LMC (0.551 $\pm$ 0.049 and 0.656 $\pm$ 0.045, respectively, with a 27 per cent AD test probability) -- the \emph{late} WC stars in M33 more closely resemble the latter in terms of mean NCR, with a 34 per cent AD test probability. The simplest explanation could be that, at a lower metallicity, a higher initial mass would be required to make a WC star through the wind mass loss channel \citep{crowther07}, which would result in higher NCRs for WC stars in the LMC at the same relative evolutionary stage as those in M33. However, removing the central regions of M33 with the highest metallicity (an ellipse with $a \sim 6$ kpc, $b \sim 3.5$ kpc and position angle 22.7 deg), and the 13 early WC stars inside this area, results in a mean NCR of 0.499 $\pm$ 0.073 at 20 Mpc for the remaining 15 early WC stars. Because of the large uncertainty, this is still consistent with the mean NCR of 0.551 $\pm$ 0.049 obtained using the whole galaxy, but does not seem to support the metallicity explanation for the $\sim$2 $\sigma$ difference between the two galaxies. We also note that, while in the LMC the NCRs of WC stars are higher than those of WN stars, as expected from progenitor mass differences and as found by \citet{leloudas10} in M83 and NGC 1313, this is not the case in M33. The results in M33 thus remain peculiar.

\subsection{Comparison with SN types}

All of the stellar samples studied here, except the sample of RSGs with log $L/L_{\odot} < 4.6$ in M33, show a mean NCR higher than that of the completely random distribution. This indicates that all these samples are correlated with the H\,$\alpha$ emission at least weakly. The mean NCR of SN impostors from H14 (0.133 $\pm$ 0.086) is consistent with a random distribution, although the errors are quite large and the median distance of the impostor sample, 8.5 Mpc, somewhat different from our simulated distances.

In order to use our analysis to set constraints on CCSN progenitors, we compare the distributions of CCSN NCRs reported in A12, K13 and H14, and their mean NCRs listed in Table \ref{table:sn_ncr}, to those of our evolved star samples at the appropriate distance. We obtain median distances for the SN samples using the redshifts (radial velocities) of the host galaxies reported in A12 and H14 and converting them into distances using $H_{0} = 70$ km s$^{-1}$ Mpc$^{-1}$, $\Omega_{m} = 0.3$, $\Omega_{\Lambda} = 0.7$. The radial velocities of the K13 sample galaxies are obtained from the NED. As the median distances of type Ic SNe in K13 and the type II-P and II-L SNe in the A12 sample are closer to 20 Mpc, we compare them to our results using the 20 Mpc images. Similarly, the median distances of types Ib, Ic (A12), IIb and IIn are closer to 35 Mpc and therefore those types are compared to our results from the 35 Mpc images. The A12 SNR is used for all these comparisons, and the star NCR distributions calculated including positional errors. We use the AD test to compare the distributions of samples containing at least ten objects. The results from these comparisons are presented in Tables \ref{table:adtest} and \ref{table:adtest_m33}, with comparisons where the mean NCRs are also consistent highlighted. In previous NCR studies such as A12 and K13, the Kolmogorov-Smirnov test was used. The AD test, however, is more sensitive to differences between the distributions, especially near the endpoints of the cumulative distribution. Comparisons between consistent distributions of stars and SNe (with consistent mean NCRs and/or with an AD probability $\ge 10$ per cent) are presented in Fig. \ref{fig:sn_vs_star}, along with LBVs versus type IIn SNe to demonstrate their difference.

The best matches for each type can be found using Tables \ref{table:main_ncr_lmc}, \ref{table:main_ncr_m33}, \ref{table:adtest} and \ref{table:adtest_m33}, defined here as both being formally consistent with the SN type according to the AD test (an AD probability $\ge 10$ per cent) and having a consistent mean NCR. These are as follows for each SN type:

\begin{figure*}
\centering
\begin{minipage}{160mm}
\begin{tabular}[t]{cc}
\includegraphics[width=8cm]{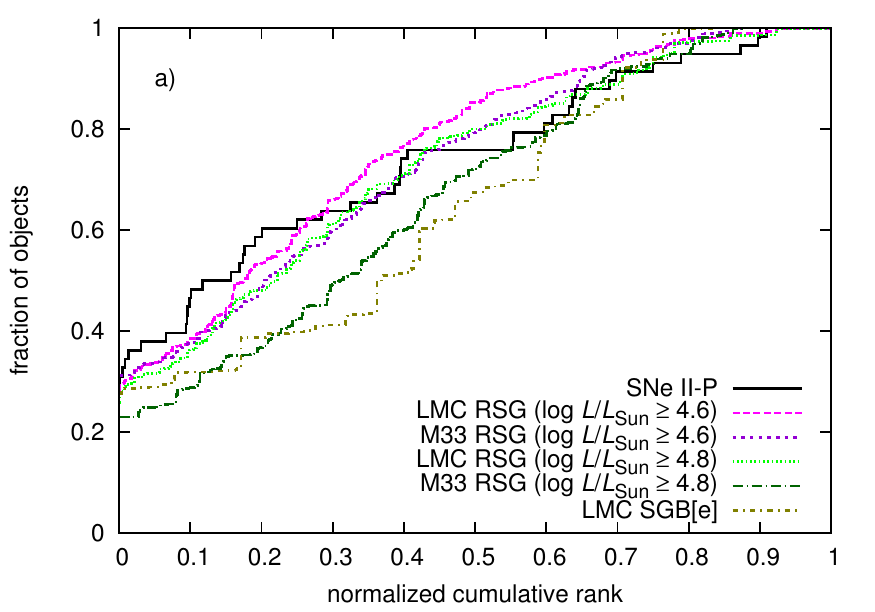} & \includegraphics[width=8cm]{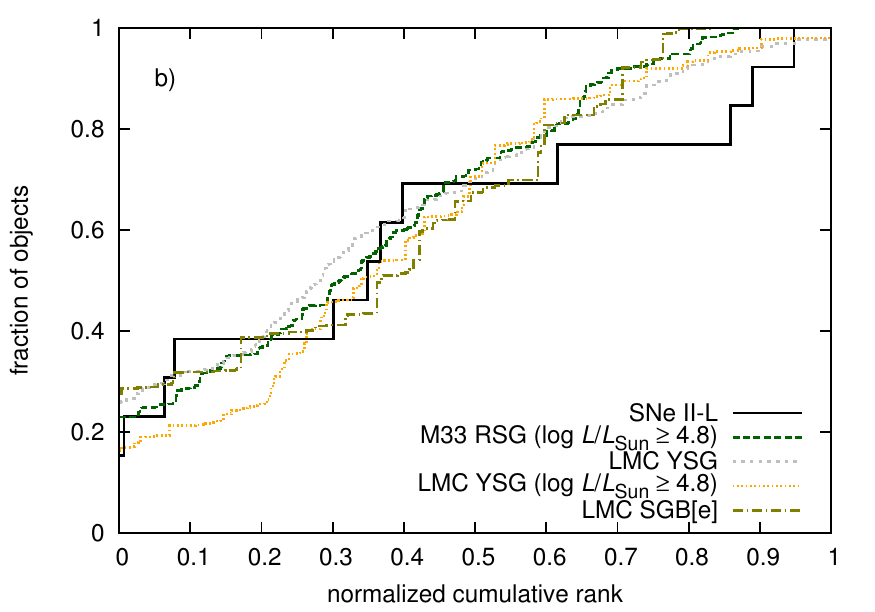} \\
\includegraphics[width=8cm]{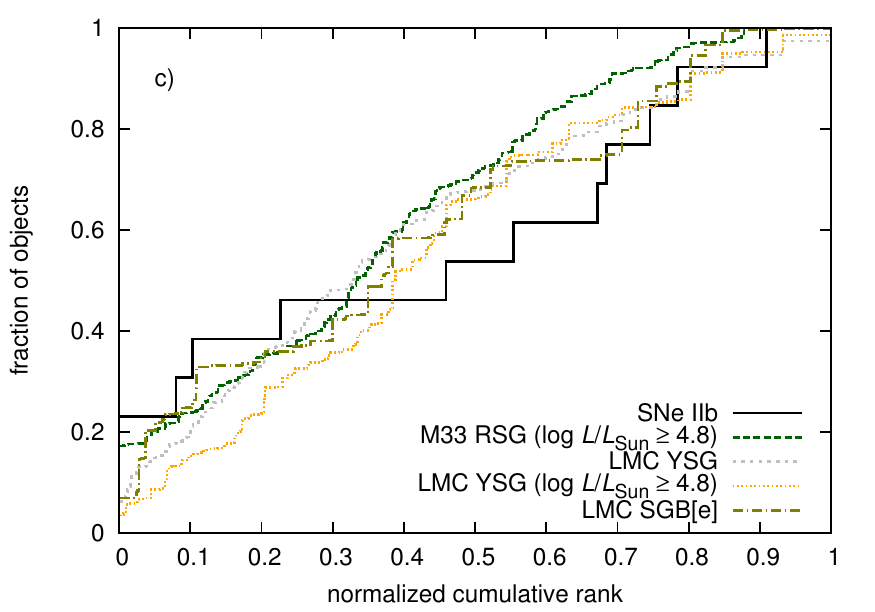} & \includegraphics[width=8cm]{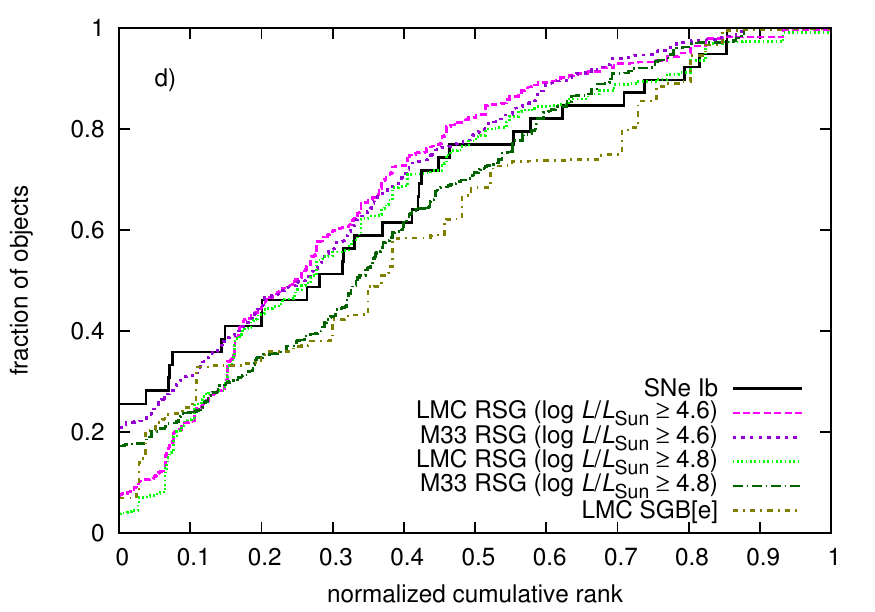} \\
\includegraphics[width=8cm]{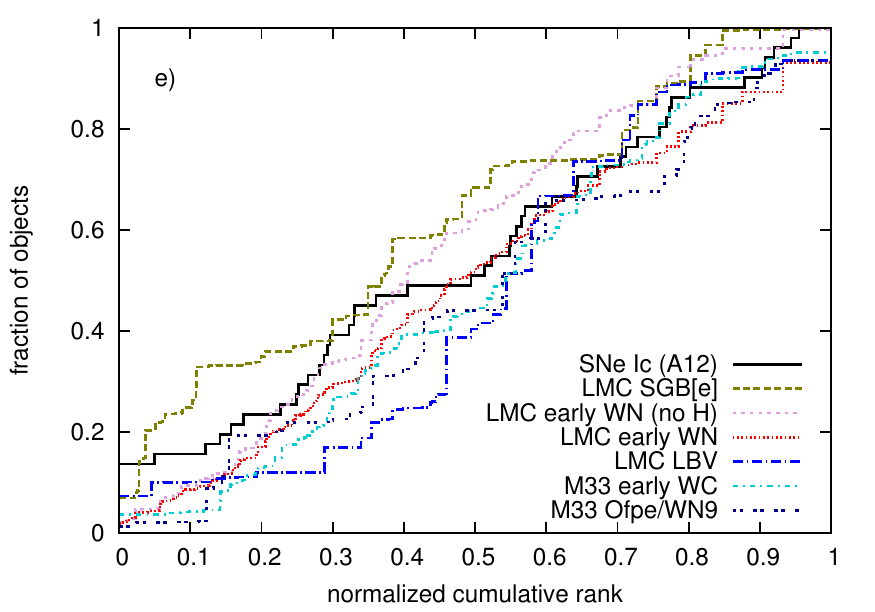} & \includegraphics[width=8cm]{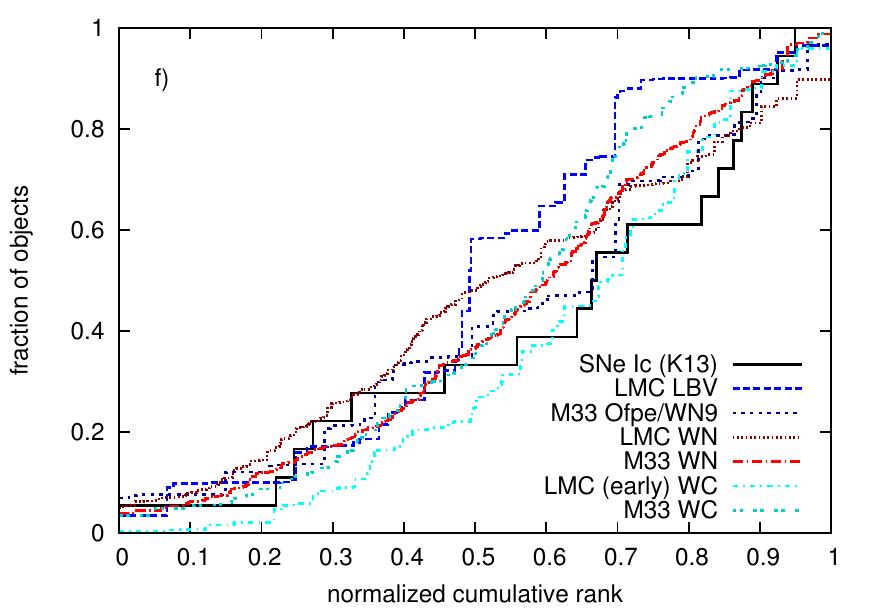} \\
\end{tabular}
\centering
\includegraphics[width=8cm]{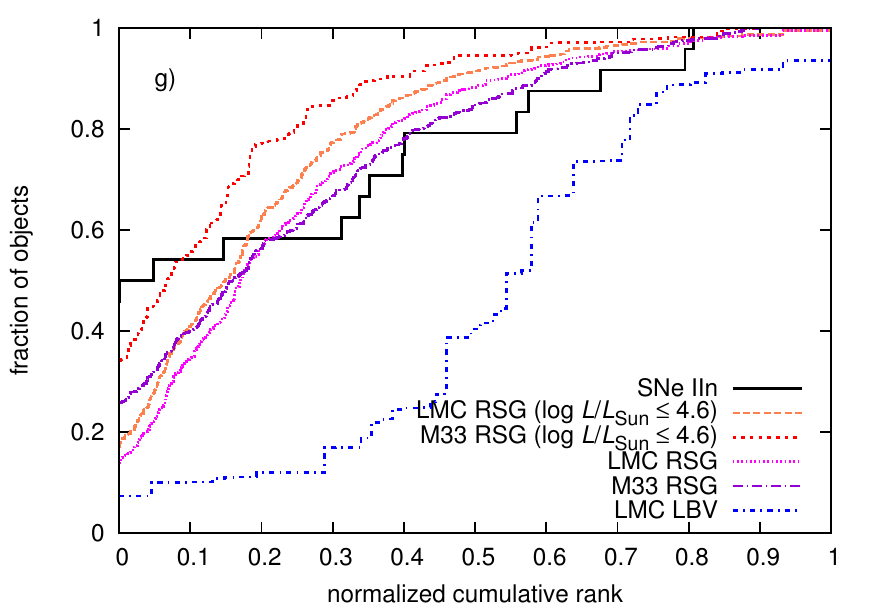}
\caption{Comparisons between the cumulative NCR distributions of various SN types and stellar samples in the LMC and/or M33. Panel a) type II-P SNe, b) type II-L, c) type IIb, d) type Ib, e) type Ic from A12, f) type Ic from K13 and g) type IIn. The stellar samples have been selected based on formal consistency according to the AD test ($\geq10$ per cent probability of the distributions being the same) and similar mean NCRs. In the case of the type IIn SNe, we also include the LBVs to demonstrate the difference of these distributions.}
\label{fig:sn_vs_star}
\end{minipage}
\end{figure*}

\begin{table*}
\centering
\begin{minipage}{116mm}
\caption{Results of the AD tests (probability in per cent that the compared distributions are the same) between the NCR distributions of SNe and stars in the LMC. All tests are performed using the results with simulated distance close to the median distance of the SNe and including positional errors. Sample comparisons where the mean NCRs are also consistent within 1 $\sigma$ are highlighted in bold.}
\label{table:adtest}
\begin{tabular}[t]{lccccccc}
    \hline
	Stellar sample & II-P & II-L & IIb & Ib & Ic (A12) & Ic (K13) & IIn \\
    \hline
	RSG & 6 & 1 & 2 & 2 & $<1$ & $<1$ & $\mathbf{<1}$\\
	~RSG (log $L/L_{\odot} < 4.6$) & $<1$ & $<1$ & $<1$ & $<1$ & $<1$ & $<1$ & $\mathbf{<1}$\\
	~RSG (log $L/L_{\odot} \geq 4.6$) & \textbf{51} & 10 & 6 & \textbf{4} & $<1$ & $<1$ & $<1$\\
	~RSG (log $L/L_{\odot} \geq 4.8$) & \textbf{72} & 29 & \textbf{12} & \textbf{4} & 1 & $<1$ & $<1$\\
	YSG & 22 & \textbf{61} & \textbf{32} & 5 & 15 & $<1$ & $<1$\\
	~YSG (log $L/L_{\odot} \geq 4.8$) & 4 & \textbf{72} & \textbf{29} & 4 & 41 & $<1$ & $<1$\\
	SG B[e] & \textbf{51} & \textbf{71} & \textbf{61} & \textbf{31} & 61 & 3 & 2\\
	LBV & $<1$ & 18 & 38 & 3 & \textbf{46} & 44 & $<1$ \\
	WN & $<1$ & 6 & 4 & $<1$ & 5 & \textbf{50} & $<1$\\
	~Early WN & $<1$ & 18 & 10 & $<1$ & \textbf{25} & 18 & $<1$\\
	~Late WN & $<1$ & $<1$ & 1 & $<1$ & $<1$ & \textbf{46} & $<1$\\
	WN (no H) & $<1$ & 19 & 11 & $<1$ & \textbf{29} & 22 & $<1$\\
	~Early WN (no H) & $<1$ & \textbf{26} & \textbf{15} & $<1$ & \textbf{33} & 1 & $<1$\\
	~Late WN (no H) & $<1$ & $<1$ & 1 & $<1$ & $<1$ & 5 & $<1$\\
	(Early) WC & $<1$ & 1 & 2 & $<1$ & 1 & \textbf{75} & $<1$\\
\hline
\end{tabular}
\end{minipage}
\end{table*}

\begin{table*}
\centering
\begin{minipage}{116mm}
\caption{As Table \ref{table:adtest}, but for M33.}
\label{table:adtest_m33}
\begin{tabular}[t]{lccccccc}
    \hline
	Stellar sample & II-P & II-L & IIb & Ib & Ic (A12) & Ic (K13) & IIn \\
    \hline
	RSG & 13 & 2 & 5 & 19 & $<1$ & $<1$ & \textbf{11}\\
	~RSG (log $L/L_{\odot} < 4.6$) & $<1$ & $<1$ & $<1$ & $<1$ & $<1$ & $<1$ & 14\\
	~RSG (log $L/L_{\odot} \geq 4.6$) & \textbf{65} & 14 & 12 & 69 & $<1$ & $<1$ & 4\\
	~RSG (log $L/L_{\odot} \geq 4.8$) & 14 & \textbf{33} & \textbf{30} & \textbf{49} & 6 & $<1$ & 1\\
	YSG & $<1$ & \textbf{32} & \textbf{17} & 2 & 21 & $<1$ & $<1$\\
	~YSG (log $L/L_{\odot} \geq 4.8$) & $<1$ & \textbf{22} & \textbf{13} & $<1$ & 27 & $<1$ & $<1$\\
	Ofpe/WN9 & $<1$ & 25 & 34 & 3 & 59 & \textbf{98} & $<1$\\
	WN & $<1$ & 1 & 3 & $<1$ & 2 & \textbf{56} & $<1$\\
	~Early WN & $<1$ & 5 & \textbf{7} & $<1$ & \textbf{20} & 3 & $<1$\\
	~Late WN & $<1$ & $<1$ & $<1$ & $<1$ & $<1$ & 22 & $<1$\\
	WC & $<1$ & 2 & 6 & $<1$ & 4 & \textbf{20} & $<1$\\
	~Early WC & $<1$ & 11 & 17 & $<1$ & 45 & 26 & $<1$\\
	~Late WC & $<1$ & $<1$ & 3 & $<1$ & $<1$ & \textbf{22} & $<1$\\
\hline
\end{tabular}
\end{minipage}
\end{table*}

\begin{itemize}
 \item \textbf{Type II-P SNe} are best matched by the RSG samples with log $L/L_{\odot} \geq 4.6$ or log $L/L_{\odot} \geq 4.8$ and SG B[e] stars. The YSGs in the LMC are consistent with type II-P according to the AD test but do not have a consistent mean NCR.
 \item \textbf{Type II-L SNe} are best matched by RSGs with log $L/L_{\odot} \geq 4.8$ in M33, YSGs (with or without the cutoff) in both galaxies, as well as early WN stars without hydrogen and SG B[e] stars in the LMC. The samples of RSGs with log $L/L_{\odot} \geq 4.6$ or log $L/L_{\odot} \geq 4.8$, LBVs and early-type WN stars are consistent with type II-L according to the AD test, but inconsistent based on the mean NCR. The small sample size of type II-L affects these results.
 \item \textbf{Type IIb SNe} are best matched by RSGs with log $L/L_{\odot} \geq 4.8$ and YSGs (with or without the cutoff) in both galaxies, as well as early WN stars without hydrogen and SG B[e] stars in the LMC. LBVs and early WN stars in the LMC, Ofpe/WN9 stars, RSGs with log $L/L_{\odot} \geq 4.6$ and early WC stars in M33 are consistent with type IIb according to the AD test but inconsistent based on the mean NCR. This is again at least partly due to the small sample size of type IIb.
 \item \textbf{Type Ib SNe} are best matched by the RSG samples with log $L/L_{\odot} \geq 4.6$ or log $L/L_{\odot} \geq 4.8$. In the LMC the AD test probabilities are lowered by having a different fraction of zero-NCR stars -- otherwise, as seen in Fig. \ref{fig:sn_vs_star}, the distributions are quite similar. SG B[e] stars also match type Ib SNe.
 \item \textbf{Type Ic SNe} (\textbf{A12}; random sample of spiral galaxies) are best matched by early WN stars (both in general and those without hydrogen) and LBVs. Several other distributions are consistent with type Ic according to the AD test but inconsistent based on the mean NCR, namely YSGs with log $L/L_{\odot} \geq 4.8$ in both galaxies, SG B[e] stars in the LMC as well as Ofpe/WN9 stars and early WC stars in M33.
 \item \textbf{Type Ic SNe} (\textbf{K13}; strongly star-forming galaxies) are best matched by Ofpe/WN9 stars in M33 and WR stars (both WN and WC) in both galaxies. The LBVs are also formally consistent with this SN sample by the AD test but inconsistent based on the mean NCR.
 \item \textbf{Type IIn SNe} are only matched by the full sample of RSGs in M33. The RSGs with log $L/L_{\odot} \leq 4.6$ are consistent with type IIn according to the AD test but inconsistent based on the mean NCR. Even these distributions seem quite different from type IIn by eye in Fig. \ref{fig:sn_vs_star}. Furthermore, the LBVs show a distribution (and a mean NCR) very different from type IIn.
\end{itemize}

\section{Systematic effects}

\begin{table*}
\centering
\begin{minipage}{140mm}
\caption{Re-calculated test NCRs in M33 after the removal of the central square kpc region from the galaxy image.}
\label{table:centralkpc}
\begin{tabular}[t]{lccccc}
    \hline
        & & \multicolumn{2}{c}{20 Mpc} & \multicolumn{2}{c}{35 Mpc} \\
	Stellar type & N & $<$NCR$>$(acc) & $<$NCR$>$(err) & $<$NCR$>$(acc) & $<$NCR$>$(err) \\
    \hline	
	RSG & 186 & 0.210 $\pm$ 0.019 & 0.207 $\pm$ 0.018 & 0.248 $\pm$ 0.018 & 0.238 $\pm$ 0.018 \\
	 ~RSG (log $L/L_{\odot} \ge 4.6$) & 118 & 0.273 $\pm$ 0.025 & 0.270 $\pm$ 0.024 & 0.306 $\pm$ 0.024 & 0.289 $\pm$ 0.023 \\
 	 ~RSG (log $L/L_{\odot} \ge 4.8$) & 69 & 0.340 $\pm$ 0.034 & 0.333 $\pm$ 0.033 & 0.376 $\pm$ 0.033 & 0.354 $\pm$ 0.032 \\
	YSG & 70 & 0.355 $\pm$ 0.029 & 0.349 $\pm$ 0.030 & 0.407 $\pm$ 0.030 & 0.395 $\pm$ 0.030 \\
	WN & 131 & 0.594 $\pm$ 0.023 & 0.567 $\pm$ 0.023 & 0.582 $\pm$ 0.023 & 0.547 $\pm$ 0.024 \\
	WC & 44 & 0.567 $\pm$ 0.037 & 0.548 $\pm$ 0.037 & 0.564 $\pm$ 0.040 & 0.537 $\pm$ 0.039 \\
\hline
\end{tabular}
\end{minipage}
\end{table*}

A general effect immediately seen in Tables \ref{table:main_ncr_lmc} and \ref{table:main_ncr_m33} is that including positional errors in the analysis results in slightly lower mean NCRs. For low initial NCRs the effect is small, and the value may even increase slightly. This is unsurprising because a small error in the position of a star in a high-NCR pixel mostly lowers its NCR, meaning that the neighbouring pixels are likely to be fainter, while a star in a low-NCR pixel can move to a fainter or brighter pixel. With $\sigma = 0.5$ arcsec and using the A12 pixel scale, $\sim$70 per cent of the stars stay in the same pixel after the error is applied, and as a result the effect of this error is not very large. There is a small increase in the effect of the positional error with distance; this is simply because the same positional error in angular coordinates corresponds to a larger error in physical coordinates as the distance of the galaxy is increased.

In addition to the steps described in Sect. 3.2 to make our images comparable to those of A12, we also investigate the effects of the removal of the central region in M33, as well as changing the distance, binning and SNR. We perform these tests on multiple subsamples with different mean NCR to test for the biases as a function of NCR. We additionally test for any spatial biases in the RSG sample in the LMC caused by the patchy coverage of the catalogue used.

In M33, we test the effect of removing the central square kiloparsec region from the H\,$\alpha$ image and excluding the stars located in this region from the stellar samples. As mentioned above, the \citet{m33sg} RSG sample is biased against this region because of crowding effects; therefore removing the region and the two RSGs located in it should make the NCR result more accurate. We also test what effect this has on some other NCRs. In short, there is no significant effect on the NCR results; the re-calculated values are listed in Table \ref{table:centralkpc}. In particular, the RSGs, which are the primary subjects of this test, show slightly increased mean NCRs, but are still well inside the errors of the original values. The YSGs, which might also have been affected by the crowding bias, show practically no change either. This confirms our earlier assertion that the spatial coverage of the M33 RSG sample is in practice unaffected by the crowding bias reported by \citet{m33sg}, assuming that the RSG population in the central region does not significantly differ from the population in the outer regions.

\begin{figure}
\centering
\includegraphics[width=\columnwidth]{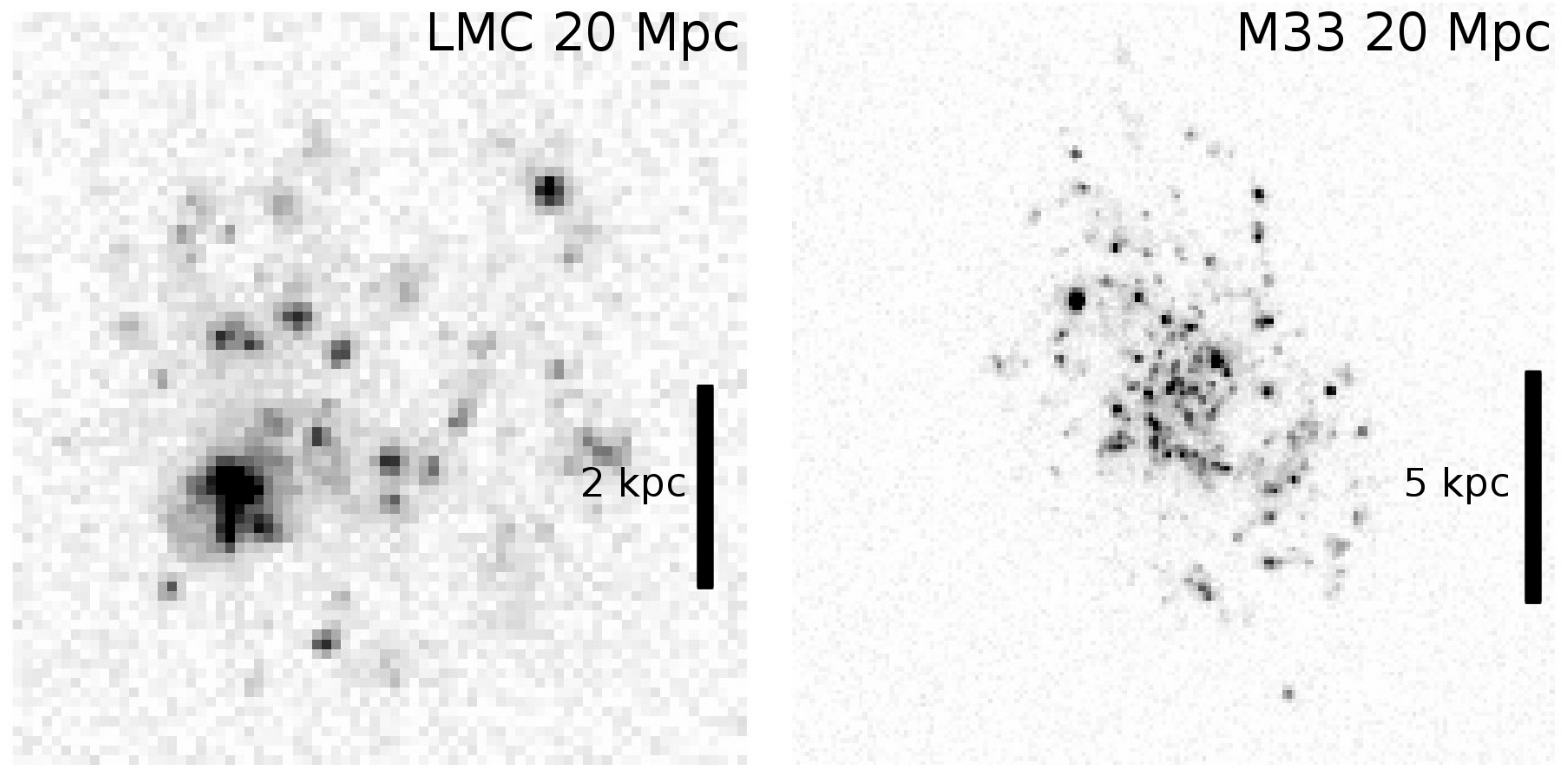}
\includegraphics[width=\columnwidth]{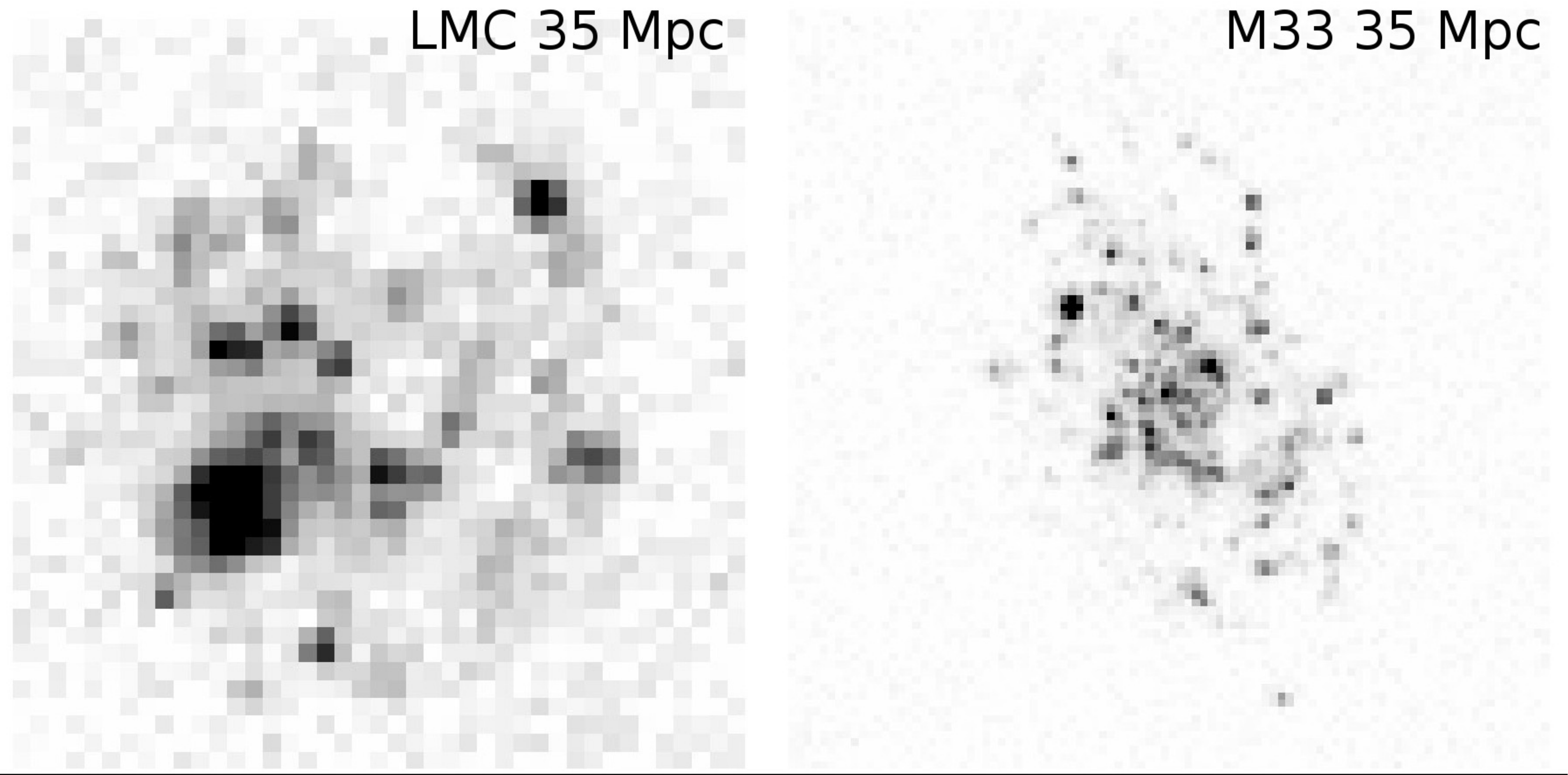}
\includegraphics[width=\columnwidth]{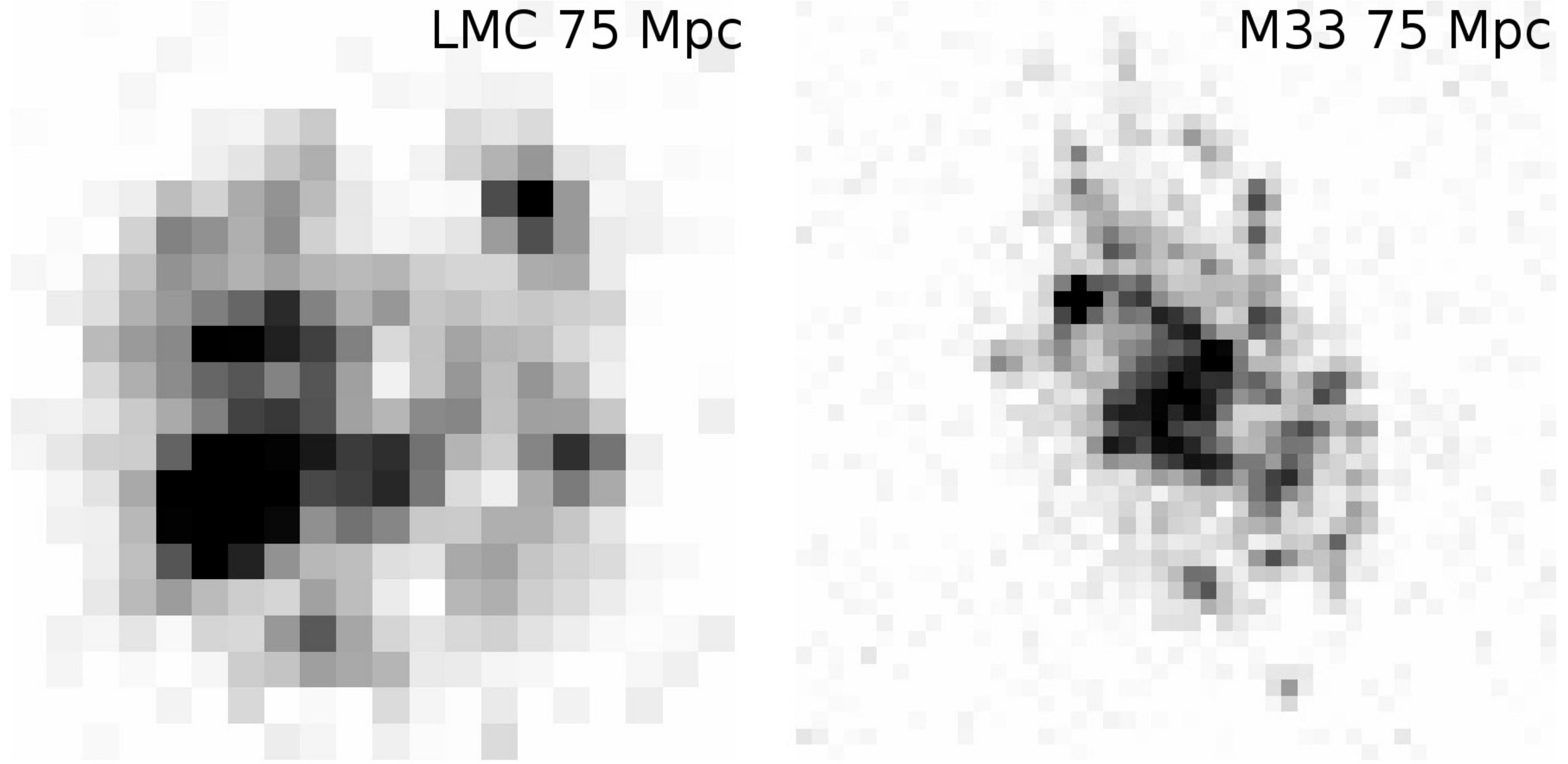}
\caption{H\,$\alpha$ images of the LMC and M33 at different simulated distances, using a $\sim$1 arcsec pix$^{-1}$ binning with a simulated distance and SNRs similar to those in A12 and K13. North is up and east is left. The black scale bar corresponds to 2 kpc in the LMC and 5 kpc in M33; the linear scale is the same in all images.}
\label{fig:ha_dists}
\end{figure}

We also calculate some test NCRs at the distance of 75 Mpc, which is at the upper end of the galaxy distances in the A12 and K13 samples, to test how distance affects the NCR determination. This is again done using the appropriate convolution -- 1 arcsec seeing at this distance corresponding to a resolution of about 360 pc, equivalent to 31.5 pix for the LMC and 44.1 pix for M33 -- and by re-binning by 31 for the LMC and by 44 for M33. To illustrate the effect on the images of LMC and M33, the convolved and re-binned H\,$\alpha$ images at different simulated distances are presented in Fig. \ref{fig:ha_dists}. Furthermore, we calculate the test NCRs for different noise levels. For each distance (20, 35 and 75 Mpc), in addition to the Poisson noise required to achieve the SNRs per pixel mentioned above, we make images with five times more and five times less noise. Tables \ref{table:dist_noise_lmc} and \ref{table:dist_noise_m33} show the effects of varying distance and noise on some NCRs in the LMC and M33, respectively, selected to cover a wide range of mean NCRs.

\begin{table*}
\centering
\begin{minipage}{130mm}
\caption{Mean NCRs ($<$NCR$>$) of selected stellar types using three different SNRs (`medium' is an SNR comparable to A12 and K13, while `high' and `low' SNRs are five times higher and five times lower, respectively) in the LMC, at simulated distances of 20, 35 and 75 Mpc. Positional errors are included in all results.}
\label{table:dist_noise_lmc}
\begin{tabular}[t]{lcccc}
    \hline
	Stellar type & N & $<$NCR$>$(high SNR) & $<$NCR$>$(medium SNR) & $<$NCR$>$(low SNR) \\
    \hline
	20 Mpc \\
    \hline
	random & 250 & 0.117 $\pm$ 0.006 & 0.095 $\pm$ 0.007 & 0.045 $\pm$ 0.006 \\	
	RSG & 543 & 0.207 $\pm$ 0.009 & 0.180 $\pm$ 0.010 & 0.106 $\pm$ 0.010 \\		
	YSG & 109 & 0.350 $\pm$ 0.027 & 0.328 $\pm$ 0.029 & 0.245 $\pm$ 0.032 \\		
	B2V & 92  & 0.487 $\pm$ 0.028 & 0.472 $\pm$ 0.030 & 0.401 $\pm$ 0.037 \\		
	B0V & 147 & 0.617 $\pm$ 0.022 & 0.610 $\pm$ 0.023 & 0.562 $\pm$ 0.028 \\		
	O5V & 13  & 0.781 $\pm$ 0.058 & 0.776 $\pm$ 0.060 & 0.750 $\pm$ 0.077 \\		
	O3V & 12  & 0.933 $\pm$ 0.029 & 0.931 $\pm$ 0.030 & 0.930 $\pm$ 0.033 \\
    \hline
	35 Mpc \\
    \hline
	random & 250 & 0.105 $\pm$ 0.006 & 0.101 $\pm$ 0.006 & 0.090 $\pm$ 0.008 \\			
	RSG & 543 & 0.232 $\pm$ 0.010 & 0.228 $\pm$ 0.010 & 0.160 $\pm$ 0.011 \\		
	YSG & 109 & 0.381 $\pm$ 0.027 & 0.375 $\pm$ 0.028 & 0.314 $\pm$ 0.032 \\		
	B2V & 92  & 0.514 $\pm$ 0.029 & 0.513 $\pm$ 0.029 & 0.463 $\pm$ 0.037 \\		
	B0V & 147 & 0.604 $\pm$ 0.022 & 0.603 $\pm$ 0.022 & 0.582 $\pm$ 0.026 \\		
	O5V & 13  & 0.751 $\pm$ 0.064 & 0.755 $\pm$ 0.063 & 0.739 $\pm$ 0.075 \\		
	O3V & 12  & 0.907 $\pm$ 0.038 & 0.911 $\pm$ 0.037 & 0.908 $\pm$ 0.039 \\
    \hline
	75 Mpc \\
    \hline
	random & 250 & 0.138 $\pm$ 0.008 & 0.140 $\pm$ 0.008 & 0.136 $\pm$ 0.009 \\		
	RSG & 543 & 0.283 $\pm$ 0.010 & 0.285 $\pm$ 0.010 & 0.244 $\pm$ 0.011 \\		
	YSG & 109 & 0.421 $\pm$ 0.029 & 0.420 $\pm$ 0.029 & 0.387 $\pm$ 0.031 \\		
	B2V & 92  & 0.550 $\pm$ 0.030 & 0.557 $\pm$ 0.030 & 0.523 $\pm$ 0.033 \\		
	B0V & 147 & 0.587 $\pm$ 0.024 & 0.587 $\pm$ 0.024 & 0.580 $\pm$ 0.025 \\		
	O5V & 13  & 0.728 $\pm$ 0.075 & 0.724 $\pm$ 0.075 & 0.723 $\pm$ 0.079 \\		
	O3V & 12  & 0.887 $\pm$ 0.048 & 0.885 $\pm$ 0.049 & 0.885 $\pm$ 0.051 \\
\hline
\end{tabular}
\end{minipage}
\end{table*}

\begin{table*}
\centering
\begin{minipage}{130mm}
\caption{As Table \ref{table:dist_noise_lmc}, but for M33.}
\label{table:dist_noise_m33}
\begin{tabular}[t]{lcccc}
    \hline
	Stellar type & N & $<$NCR$>$(high SNR) & $<$NCR$>$(medium SNR) & $<$NCR$>$(low SNR) \\
    \hline
	20 Mpc \\
    \hline
	random & 250 & 0.104 $\pm$ 0.007 & 0.089 $\pm$ 0.007 & 0.050 $\pm$ 0.007 \\		
	RSG & 188 & 0.221 $\pm$ 0.017 & 0.201 $\pm$ 0.018 & 0.122 $\pm$ 0.018 \\
	YSG & 74  & 0.371 $\pm$ 0.028 & 0.353 $\pm$ 0.030 & 0.225 $\pm$ 0.034 \\		
	WN  & 139 & 0.572 $\pm$ 0.021 & 0.566 $\pm$ 0.022 & 0.481 $\pm$ 0.029 \\		
    \hline
	35 Mpc \\
    \hline
	random & 250 & 0.117 $\pm$ 0.007 & 0.111 $\pm$ 0.008 & 0.069 $\pm$ 0.008 \\	
	RSG & 188 & 0.233 $\pm$ 0.017 & 0.228 $\pm$ 0.017 & 0.176 $\pm$ 0.019 \\		
	YSG & 74  & 0.399 $\pm$ 0.029 & 0.399 $\pm$ 0.030 & 0.328 $\pm$ 0.037 \\		
	WN  & 139 & 0.547 $\pm$ 0.023 & 0.551 $\pm$ 0.023 & 0.509 $\pm$ 0.027 \\		
    \hline
	75 Mpc \\
    \hline
	random & 250 & 0.140 $\pm$ 0.008 & 0.139 $\pm$ 0.008 & 0.108 $\pm$ 0.009 \\	
	RSG & 188 & 0.251 $\pm$ 0.017 & 0.250 $\pm$ 0.017 & 0.217 $\pm$ 0.018 \\		
	YSG & 74  & 0.452 $\pm$ 0.031 & 0.455 $\pm$ 0.031 & 0.433 $\pm$ 0.035 \\		
	WN  & 139 & 0.517 $\pm$ 0.024 & 0.520 $\pm$ 0.024 & 0.494 $\pm$ 0.026 \\		
\hline
\end{tabular}
\end{minipage}
\end{table*}

\begin{table*}
\centering
\begin{minipage}{111mm}
\caption{Mean NCRs ($<$NCR$>$) of selected stellar types at the simulated distance of 35 Mpc in the LMC, calculated using three different pixel scales: $\sim$1 arcsec pix$^{-1}$ (A12), $\sim$0.38 arcsec pix$^{-1}$ (K13) and $\sim$0.19 arcsec pix$^{-1}$ (the pixel scale of the ALFOSC instrument). Positional errors are included in all results.}
\label{table:bin_lmc}
\begin{tabular}[t]{lcccc}
    \hline
	Stellar type & N & $<$NCR$>$(A12) & $<$NCR$>$(K13) & $<$NCR$>$(ALFOSC) \\
    \hline
	random & 250 & 0.101 $\pm$ 0.006 & 0.126 $\pm$ 0.008 & 0.123 $\pm$ 0.007 \\	
	RSG & 543 & 0.228 $\pm$ 0.010 & 0.207 $\pm$ 0.010 & 0.200 $\pm$ 0.010 \\		
	YSG & 109 & 0.375 $\pm$ 0.028 & 0.358 $\pm$ 0.028 & 0.349 $\pm$ 0.028 \\		
	B2V & 92  & 0.513 $\pm$ 0.029 & 0.489 $\pm$ 0.030 & 0.486 $\pm$ 0.030 \\		
	B0V & 147 & 0.603 $\pm$ 0.022 & 0.596 $\pm$ 0.022 & 0.597 $\pm$ 0.022 \\		
	O5V & 13  & 0.755 $\pm$ 0.063 & 0.747 $\pm$ 0.062 & 0.745 $\pm$ 0.063 \\		
	O3V & 12  & 0.911 $\pm$ 0.037 & 0.897 $\pm$ 0.035 & 0.894 $\pm$ 0.035 \\

\hline
\end{tabular}
\end{minipage}
\end{table*}

\begin{table*}
\centering
\begin{minipage}{111mm}
\caption{As Table \ref{table:bin_lmc}, but for M33.}
\label{table:bin_m33}
\begin{tabular}[t]{lcccc}
    \hline
	Stellar type & N & $<$NCR$>$(A12) & $<$NCR$>$(K13) & $<$NCR$>$(ALFOSC) \\
    \hline
	random & 250 & 0.111 $\pm$ 0.008 & 0.096 $\pm$ 0.007 & 0.105 $\pm$ 0.009 \\		
	RSG & 188 & 0.228 $\pm$ 0.017 & 0.208 $\pm$ 0.018 & 0.205 $\pm$ 0.018 \\
	YSG & 74  & 0.399 $\pm$ 0.030 & 0.368 $\pm$ 0.031 & 0.365 $\pm$ 0.030 \\		
	WN  & 139 & 0.551 $\pm$ 0.023 & 0.550 $\pm$ 0.023 & 0.552 $\pm$ 0.023 \\
\hline
\end{tabular}
\end{minipage}
\end{table*}

\begin{figure}
\centering
\includegraphics[width=\columnwidth]{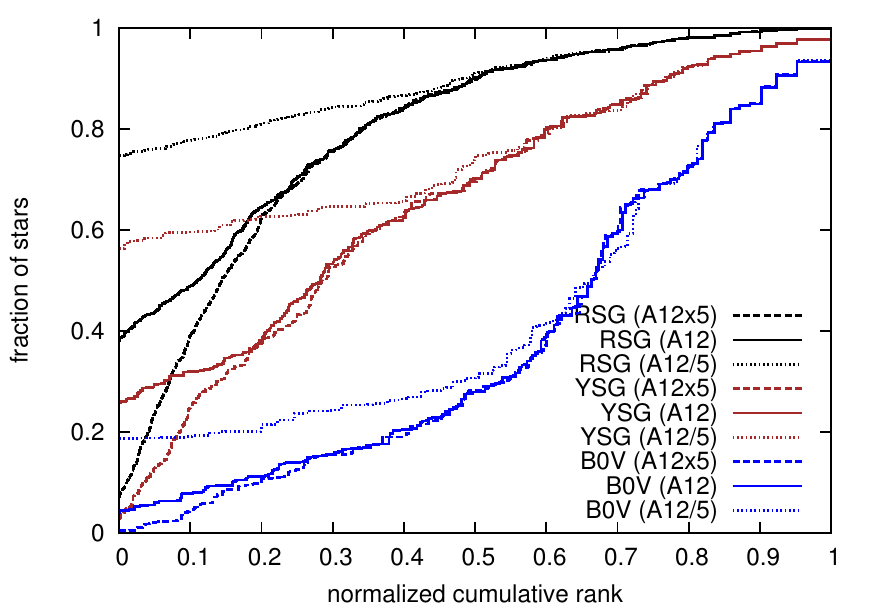}
\caption{Illustration of the effect of noise in the H\,$\alpha$ image on the NCR distribution. The distributions with the dashed lines were obtained using an SNR approximately five times better than that of the A12 images; the solid lines represent the A12 SNR; the dotted lines represent a five times worse SNR. The main effect is an increase in the number of zero-NCR stars, and this effect is most significant on initially low NCRs.}
\label{fig:noise_ncr}
\end{figure}

As can be seen from Tables \ref{table:dist_noise_lmc} and \ref{table:dist_noise_m33}, increasing the distance generally shifts NCRs closer to $\sim$0.6 in the LMC and $\sim$0.5 in M33, that is, mean values under $\sim$0.6 or $\sim$0.5 are increased with increasing distance and values over these are decreased. The effect is significant for several samples, and demonstrates the need to simulate the appropriate distance before comparing our results to those of SNe. As the distance increases, the spatial resolution worsens, and sharp features are smeared over a larger apparent area. After the convolution, an initially low-NCR star may end up in a pixel with higher flux, while an initially very high NCR does not change much, as the brightest pixels remain so even with the convolution. Adding noise (and thus lowering the SNR) generally has a very different effect: initially low NCRs are lowered further. High values are also lowered, but they change much less because as the noise level increases, more pixels become indistinguishable from the background and their NCR becomes zero. This affects faint pixels first, resulting in more zero NCRs in initially low-NCR samples. Fig. \ref{fig:noise_ncr} illustrates this effect on the cumulative distributions of the NCRs of some stellar types. The NCRs of pixels that are relatively bright but not the very brightest also decrease: less and less flux originates in the pixels fainter than them, while the bright pixels themselves are comparatively unaffected by the noise. The difference between `high' and `medium' (A12) SNR is small above NCR $\sim$0.15, indicating that not much H\,$\alpha$ emission is lost at the A12 noise levels and that the SNR in the A12 study is adequate. With further degradation of the SNR, however, the effect on the NCR becomes significant up to NCR $\sim$0.4.

Furthermore, we calculate some test NCRs with different binning. The NCRs of K13 were calculated using images with 0.38 arcsec pix$^{-1}$ and, in the case of type Ic SNe, are somewhat different from those in A12. To investigate what bias, if any, the different binning may introduce, we re-bin the 35 Mpc image (close to the median distance of both samples) to have this pixel scale. This means re-binning the original LMC image by 6 and that of M33 by 8. We also compare these to images with the pixel scale of the unbinned ALFOSC H\,$\alpha$ images of K13, 0.19 arcsec pix$^{-1}$ (binning by 3 for the LMC and by 4 for M33). The effect of increasingly coarse binning without convolution, as shown in Tables \ref{table:bin_lmc} and \ref{table:bin_m33}, is qualitatively the same as that of increased distance, which includes both convolution and re-binning. More and more pixels are averaged into one, and an initially low-NCR star can end up in a higher-NCR pixel. Concomitantly, high NCRs can be `diluted' by the inclusion of faint pixels in the same bin, but the highest values generally remain so. This effect is, however, not significant for most samples.

Finally, we test whether the spatial coverage of the \citet{lmcsg}, with 64 fields spread seemingly randomly over the visible extent of the LMC and with gaps in-between, introduces a spatial bias in the NCR analysis. Conceivably, the regions between the observed fields might host preferentially low or high numbers of RSGs. Using the 2MASS catalogue, the Naval Observatory Merged Astrometric Dataset\footnote{http://www.usno.navy.mil/USNO/astrometry/optical-IR-prod/nomad} (NOMAD) and the Wide-field Infrared Survey Explorer catalogue\footnote{http://wise.ssl.berkeley.edu/} (WISE), we search for RSG candidate stars that fulfil the following criteria:
\begin{itemize}
\item the proper motion of the star, obtained from NOMAD, is compatible with the Magellanic Clouds \citep{gfernandez};
\item $K_{\mathrm{2MASS}} \leq 10.2$ mag, the magnitude limit for RSGs in the \citet{lmcsg} catalogue, corresponding to an absolute magnitude of $M_{K} = -8.3$ mag;
\item the star is visible in the WISE 3.4 and 4.6 $\mu$m bands.
\end{itemize}
Roughly three times more 2MASS stars in the LMC (1670) match these criteria than are included in the \citet{lmcsg} RSGs (543), which agrees with the completeness estimate by the catalogue authors. A visual comparison between the spatial distributions of these candidates and the RSG sample shows no obvious bias toward any particular surface brightness. The mean NCR for the candidates, using the image of the LMC at 20 Mpc with no positional errors, is 0.188 $\pm$ 0.006, which is consistent with the mean NCR of the RSG sample at that distance (0.182 $\pm$ 0.010). Furthermore, an AD test between the two distributions reveals that they are formally consistent (18 per cent probability of a common population). Thus we conclude that no significant biases are introduced by the coverage of the \citet{lmcsg} catalogue.

\section{Discussion}

Looking at the results of main-sequence stars in Tables \ref{table:main_ncr_lmc} and \ref{table:main_ncr_m33}, with or without including positional errors, one can easily see a trend of higher NCR with earlier spectral type and thus higher initial mass. The trend is illustrated in Fig.~\ref{fig:star_ncr}. This sequence of mean NCRs provides evidence that NCR is a useful indicator of progenitor lifetime and mass. Unfortunately, a direct comparison between the NCRs of main-sequence stars and SNe is not feasible. SNe are by definition at the terminal stage, while the main-sequence stars in a galaxy are, on average, at the midpoint of their life assuming a constant star formation rate in the galaxy over the past few tens of Myr. Therefore an average OB main-sequence star still has time to move during its lifetime and change its NCR. Furthermore, the coordinates and spectral types of these stars are from SIMBAD and thus from multiple independent studies that may have concentrated on specific locations such as 30 Doradus. Therefore there may be severe completeness biases in the samples of main-sequence stars reported in SIMBAD. A sign of a bias is visible by eye in Fig.~\ref{fig:samples2}, where many main-sequence B stars are concentrated in a (marked) rectangular area with no strong H\,$\alpha$ emission sources. The NCRs of main-sequence stars should only be regarded as qualitative evidence of a lifetime-NCR relation.

More fruitful and quantitative results can be obtained using the spatially unbiased, or even complete, samples of evolved stars. This is not only because of the biases mentioned, but because the time a star spends as a supergiant, LBV or WR star after the main sequence is short. The longest of these phases, the RSG phase, lasts up to about 1 Myr. Thus at a (projected) velocity of $\sim$10 km s$^{-1}$, normal for an RSG in the LMC \citep{om07}, a star can move about 10 pc -- roughly half a pixel in our simulated 20 Mpc images -- during this phase. The YSG and LBV phases last on the order of 10$^{4}$ or 10$^{5}$ yr in single star evolution models \citep{lmcsg,smith14}, while the WR phase lasts a few times 10$^{5}$ yr \citep{crowther07}. Therefore, unlike during the main-sequence phase, a star does not move much during these phases and the H~{\sc ii} region similarly does not change much, and hence its NCR should be roughly the same as that of the resulting SN. If the YSG phase does not end in a SN \citep[doing so requires rotation and enhanced mass loss according to][]{georgy12_2}, it will be followed by another evolutionary stage, but this stage should also be short.

There is immediate evidence for the validity of the NCR method in the results using evolved stars as well. The initial masses of both LBVs and WN stars are in the $\geq 25 M_{\odot}$ range \citep[][respectively, including the low-luminosity LBVs]{smith04,crowther07}, and the mean NCRs of LBVs and WN stars in general are consistent within the errors. WC stars have a higher mean NCR because of their higher minimum initial mass of $\sim 40 M_{\odot}$ \citep{crowther07}. Even though WC stars are at a relatively later stage of evolution than WN stars, the higher mass range means they are still younger on average. The NCRs of these massive evolved stars are significantly higher than those of the mostly lower-mass YSGs and RSGs. Inside the RSG and YSG samples, a higher luminosity cutoff (and hence higher minimum mass) results in a higher NCR as well. Furthermore, SG B[e] stars have a lower NCR than LBVs, which is compatible with the \citet{smith15} argument that they are lower-mass analogs of LBVs. The pixel statistics method was criticized by \citet{crowther13} and \citet{smith15}, who argued that, as the brightest giant H~{\sc ii} complexes are the longest-lived ones, an association with them should not be a measure of lifetime, and that the most massive stars should be located in small individual H~{\sc ii} regions. However, as discussed in Sect. 3, we would still expect a lower likelihood for a longer-lived star to remain in its native giant H~{\sc ii} region (into which the smaller regions would blend at the spatial scales probed by A12), and our NCR results indicate that this is true.

In addition to connecting SNe to their progenitor stars, the mean NCRs of the subsamples of evolved and main-sequence stars can, in principle, constrain stellar evolutionary channels. If a sample of evolved stars has a higher mean NCR than one of main-sequence stars, stars in the latter sample cannot predominantly be the precursors of the former -- even though some exceptions to this rule would be possible. However, there are problems with this approach. As the NCR of a main-sequence star can change significantly before the following stages of its evolution, a direct comparison between the NCR distributions is no more fruitful than between the distributions of main-sequence stars and SNe. Furthermore, an unbiased catalogue of main-sequence stars would be required for meaningful constraints. One can, nonetheless, use this approach as another consistency check on the NCR method. Some mean NCRs of main-sequence stars in the LMC do seem incompatible with some of the evolved stars: types B1V and B2V (1-$\sigma$ upper limit of 0.554 at 20 Mpc without positional errors) vs. WC stars ($\ge 0.611$) or late WN stars ($\ge 0.618$); types B0V and later ($\le 0.649$) vs. classical LBVs ($\ge 0.659$). With other evolved types, such constraints cannot be made, as their mean NCRs are consistent with or lower than those of any of our main-sequence subsamples. As the precursors of WR stars and LVBs are believed to be much more massive than B-type main-sequence stars, these results are consistent with our present view of stellar evolution.

\subsection{SN progenitor masses}

As the lifetime-NCR connection now seems secure, we can try to use it to constrain the SN progenitors. Our first star-SN comparison is between type II-P SNe and RSGs, as the latter are already firmly established as the progenitors of the former. A luminosity of log $L/L_{\odot} = 4.6$ corresponds to an initial mass of $\sim$9 $M_{\odot}$ for an RSG with a weak dependence on metallicity according to \citet{smartt+09}, who use the Cambridge University {\sc stars} code \citep{starscode}. This is close to the minimum initial mass required for a type II-P progenitor \citep[e.g.][]{smartt09,smartt15}. RSGs with this luminosity cutoff provide a good match to type II-P SNe in both galaxies (Sect. 4.2). The comparison is affected by lifetime effects. While the sample of type II-P SNe, with detected SNe covering the entire sky, probes the entire initial mass function (IMF) of type II-P progenitors, the probability of a star to be observed as an RSG depends on the corresponding RSG lifetime. Therefore, the RSG samples include fewer high-mass RSGs with high NCRs than expected from the IMF, and the RSGs with log $L/L_{\odot} \geq 4.8$ are also consistent with type II-P SNe in the LMC. In M33, the more top-heavy RSG distribution in the catalogue (because of completeness issues) serves to offset this. As such, we conclude that the type II-P and RSG (log $L/L_{\odot} \ge 4.6$) distributions are consistent. Thus the NCR method succeeds in reproducing a progenitor range for type II-P SNe similar to that predicted by stellar evolution models and confirmed by direct progenitor detections.

The A12 IIb sample size is small and it is difficult to narrow down the possible progenitors. However, the similarity between our YSG samples and the type IIb SNe is consistent with the detected supergiant progenitors of type IIb SNe and with the 12 -- 16 $M_{\odot}$ initial mass range suggested by \citet{jerkstrand15}. The initial mass range of YSGs in the catalogues \citep{m33sg,lmcsg} is between about 10 and 25 $M_{\odot}$ according to Geneva stellar evolution models, depending on rotation. The average luminosity of the YSGs in the LMC catalogue is log $L/L_{\odot} = 4.7$, corresponding to about 15 $M_{\odot}$, while in M33 the higher completeness limit results in an average of log $L/L_{\odot} = 5.0$, corresponding to between 15 and 20 $M_{\odot}$. Type II-L SNe are consistent with similar stellar samples and have a mean NCR consistent with type IIb SNe, indicating a similarity in progenitor mass as well. Other results \citep[e.g.][]{2009hd, faran14, terreran16} are also consistent with masses around 15 $M_{\odot}$ for type II-L progenitors. The small sample size of type II-L, again, makes it difficult to draw robust conclusions. However, the presence of a binary companion may result in stronger mass loss in type IIb progenitors despite the similar mass, leaving behind less hydrogen than would be expected from a single star.

As for type Ib and Ic SNe, A12 and K13 found their NCR distributions to differ significantly. The indication of this is that while type Ic progenitors may be relatively massive (whether single or binary), type Ib progenitors are not, on average, much more massive than the RSG progenitors of type II-P and therefore are likely to be in interacting binaries. As Fig. \ref{fig:sn_vs_star} shows, and as is apparent from Tables \ref{table:main_ncr_lmc}, \ref{table:main_ncr_m33}, \ref{table:adtest} and \ref{table:adtest_m33}, type Ib SNe are consistent with RSGs with a log $L/L_{\odot} \geq 4.6$ or log $L/L_{\odot} \geq 4.8$ cutoff. While RSGs themselves cannot be the immediate progenitors of stripped-envelope SNe such as type Ib, this does provide an indication that the initial masses of type Ib progenitors are in a range similar to that of type II-P progenitors, located in interacting binaries. This is consistent with the suggested binary progenitor of iPTF13bvn, which had an initial mass of 10 -- 12 $M_{\odot}$ \citep{em16}, similar to those of the RSG progenitors of several type II-P SNe \citep{smartt15}. This result is also corroborated by, for example, \citet{hanin_a} who find type Ib progenitors to be older than those of type Ic and likely to be located in interacting binaries. The slightly higher mean NCR compared to type II-P is likely a result of the higher median distance of the Ib sample (Sect. 5).

Type Ic SNe, on the other hand, have an NCR distribution, and a mean NCR, similar to the early-type WN stars. The initial mass range of single WN stars is $\gtrsim 25 M_{\odot}$ \citep{crowther07} at $Z_{\odot}$, and \citet{lmcwr} find a limit of $\gtrsim 20 M_{\odot}$ in the LMC; the similarity of this stellar type with type Ic SNe lends support to type Ic progenitors sharing this initial mass limit. Whether type Ic SNe are actually the end-points of single WN evolution (possibly through a subsequent short WC phase) or originate in more massive interacting binaries than type Ib cannot be said based on our results. A mix of WC stars and lower-mass interacting binaries is also possible. The fact that no type Ic SN progenitor has yet been detected is sometimes used to exclude WR stars as a significant type Ic SN progenitor channel \citep[e.g.][]{smith14}. However, \citet{yoon12} and \citet{groh13} argued that these stars would be optically faint shortly before the SN despite their high bolometric luminosity, and the lack of detected progenitors is thus not surprising. The K13 sample of type Ic SNe has an even higher mean NCR and a distribution more similar to WC or Ofpe/WN9 stars or a mixture of early- and late-type WN stars. The difference between the NCR distributions in A12 and K13 is likely a combination of factors: a different median distance and binning in the two studies contributes to the higher NCRs in K13, but a selection effect or a physical difference between the samples of host galaxies is probably needed as well. The K13 galaxies were selected for a high far-infrared luminosity and thus a high star formation rate. It is possible that a recent starburst episode in these galaxies results in type Ic progenitors being younger (and thus more massive, perhaps WC stars) on average -- or still being preferentially located inside their native H~{\sc ii} region -- thus artificially increasing their NCR. Differences between the NCR distributions of other CCSN types between A12 and K13 are not significant; longer-lived progenitors may not be similarly affected by a very recent starburst episode.

LBVs have been suggested by various studies as the immediate progenitors of at least some type IIn SNe \citep[][]{trundle08, galyam09, smith11b, vandyk12, fransson14}. The initial masses of these progenitors are estimated to be above 30 or even 50 $M_{\odot}$ \citep{galyam09, smith11b}. It is, however, clear from Fig. \ref{fig:sn_vs_star} that the NCR distributions of type IIn SNe and LBVs do not match. We also note that some lower-luminosity LBVs are actually post-RSG stars according to the models of \citet{meynet05} and \citet{ekstrom12} and should be treated separately. The three classical LBVs in the LMC are consistent with stars of types O8V through O4V, a result similar to that of \citet{antismith} (although the numbers of these objects in both studies are very small). Lower-luminosity LBVs, on the other hand, are consistent with YSGs with log $L/L_{\odot} \geq 4.8$. Both of these types have a mean NCR significantly higher than that of type IIn SNe; therefore only a small fraction of IIn progenitors can be LBVs. Type IIn is a very diverse class of transients, likely to include SNe from multiple, very different channels such as super-AGB stars exploding as electron capture SNe \citep[as suggested for 1994W-like SNe by e.g.][]{kankare12}, extreme RSGs \citep[as suggested for 1998S-like SNe by e.g.][]{taddia15} or type Ia events disguised as IIn \citep[e.g.][]{iacsm1}. A combination of the latter three types, with a few LBVs added into the mix, could account for the observed NCR distribution. For example, RSGs with log $L/L_{\odot} < 4.5$, corresponding to $M \lesssim 8 M_{\odot}$ \citep{smartt+09}, at 35 Mpc in the LMC have a mean NCR of 0.143 $\pm$ 0.012, which is also close to the NCR of type Ia SNe (A12). A population of 70 per cent such RSGs, which probably include a significant fraction of AGB stars, and 30 per cent LBVs drawn from both LBV subclasses, results in a mean NCR of 0.258 $\pm$ 0.026. This is still consistent with the type IIn SNe within the errors.

SN impostor events have been connected to LBVs \citep[e.g.][]{vandyk02,maund06, tartaglia15}; however, there is evidence \citep[][and references therein]{smith15} that lower-mass stars could also be responsible for similar events \citep[e.g. a 12 -- 17 $M_{\odot}$ star suggested to be the progenitor of NGC300 OT2008-1 by][]{gogarten09}. These events were investigated along with type IIn SNe by H14, but their mean NCR is only 0.133 $\pm$ 0.086 -- although the median distance of their sample is only 8.5 Mpc and, as we have seen, increasing the distance will also increase low NCRs. One also needs to keep in mind that as SN impostors are significantly fainter than bona fide SNe, their detection may be heavily biased against bright backgrounds, and such regions often have high NCRs (this also causes the low median distance of the sample). As such, the NCR of the impostors may be heavily biased toward low values.

\subsection{Effects of binary evolution}

The initial mass ranges of WR stars reported by \citet{crowther07} and \citet{lmcwr}, the evolutionary tracks presented in \citet{m33wr} and \citet{lmcwr} and the initial masses of RSG progenitors of CCSNe reported by e.g. \citet{smartt+09} are the results of single star models. However, binary interaction dominates the evolution of massive stars: over 70 per cent of all O-type stars are expected to exchange mass with a companion, with a binary merger in a third of the cases \citep{sana12}. Only about half of apparently single massive stars, those with an amplitude of radial velocity variation less than 10 km s$^{-1}$, are truly single \citep{demink14}, with the other half being wide binaries before mass transfer or products of mass transfer or a merger. Therefore it is worthwhile to discuss possible effects that the multiplicity of a star may have on the NCR-mass connection.

The time a star spends as an RSG in a binary system is shortened by the mass transfer by a factor of two or three (Eldridge, Izzard \& Tout 2008). Thus the binary RSGs that are included in the sample make the average duration of the RSG phase shorter. Therefore the NCRs of RSGs in binary systems should be closer to those of the resulting SNe than expected from single stars. However, as the phase is already short compared to the total lifetime of the star, the change in the mean NCR should not exceed our 1 $\sigma$ uncertainty.

The apparent binary fraction in the \citet{lmcwr} catalog is about 15 per cent, with an additional 20 per cent labelled `binary suspects'. However, based on the statistics reported by \citet{demink14}, the total fraction of WN stars affected by binary evolution may be as high as about 60 per cent. \citet{eit08} found that while the change in the lifetime of a WR star of a particular mass is minor between single- and binary-star models, the minimum initial mass for a WR star in a multiple system decreases to around 15 $M_{\odot}$. In addition, the mean NCR may be changed by effects of mergers and mass transfer, which in some cases lengthen the lifespan of the star beyond what is expected from a single star of the same initial mass \citep{demink14}. Because of the uncertainties involved in the mass-lifetime connection in binary systems, it is difficult to say what the effect on the NCRs is. This convolution is, however, already included in the NCRs of a particular SN or stellar sample. The average initial masses of the WR stars may in any case be lowered by the effect of binaries, which would also impact the inferred progenitor mass range of type Ic SNe.

\section{Conclusions}

We have used the pixel statistics method to calculate the spatial correlations of massive stars and H\,$\alpha$ emission in two nearby galaxies, the LMC and M33. We have simulated the typical distances, noise levels and binning of previous pixel statistics studies of more distant SN host galaxies and, for the first time, directly compared the NCR distributions of different types of SNe to different stellar samples. We find the following:

   \begin{itemize}
      \item The pixel statistics method shows a correlation between a shorter lifetime, and hence higher initial mass, and higher NCR. This applies to both the main sequence OB stars in the LMC and evolved stars in both galaxies. In addition, the NCRs of the stellar samples are generally consistent between the two galaxies despite different metallicities, structures and star formation histories. Thus we find NCR to be a valuable tool in studying the progenitor stars of CCSNe in the local universe indirectly.
      \item The results from the pixel statistics method do not change significantly between an SNR typical to A12 and one five times higher. Whether the pixel scale is typical to A12 or five times smaller likewise does not have a significant effect. Therefore the aforementioned studies do not contain obvious biases resulting from these factors.
      \item The NCR distribution of RSGs with initial masses $\gtrsim$9 $M_{\odot}$ matches that of type II-P SNe, consistently with the II-P progenitor masses derived from direct progenitor detections. Similarly, the NCRs of YSGs with an average initial mass of $\sim$15 $M_{\odot}$ are consistent with those of type IIb SNe. Type II-L SNe are consistent with sharing a progenitor mass range with type IIb, suggesting another factor such as the presence or absence of a binary companion as the main difference between these types.
      \item The NCR distribution of type Ib SNe is consistent with stars with initial masses $\gtrsim$9 $M_{\odot}$, indicating that the dominant progenitors for these SNe are not single WR stars but lower-mass interacting binaries. The NCR distribution of type Ic SNe, on the other hand, is consistent with early-type WN stars, typically with initial masses $\gtrsim$20 $M_{\odot}$ according to single-star evolution models, with a contribution from binaries down to 15 $M_{\odot}$.
      \item Both classical (initially $\gtrsim$50 $M_{\odot}$) and low-luminosity (initially $\gtrsim$25 $M_{\odot}$) LBV stars have an NCR distribution inconsistent with type IIn SNe, with a significantly higher mean NCR. Therefore, the progenitors of type IIn SNe are probably mostly stars with a lower initial mass.
   \end{itemize}

\begin{acknowledgements}

We thank the anonymous referee for many good suggestions that helped improve the paper, and Philip Massey and Brian Skiff for advice on the stellar catalogues in the LMC.

This research has made use of the Southern H-Alpha Sky Survey Atlas (SHASSA), which is supported by the National Science Foundation; the SIMBAD database, operated at CDS, Strasbourg, France; the NASA/IPAC Extragalactic Database (NED) which is operated by the Jet Propulsion Laboratory, California Institute of Technology, under contract with the National Aeronautics and Space Administration; and data products from the Wide-field Infrared Survey Explorer, which is a joint project of the University of California, Los Angeles, and the Jet Propulsion Laboratory/California Institute of Technology, funded by the National Aeronautics and Space Administration.

This work was partly supported by the European Union FP7 programme through ERC grant number 320360. RGI thanks the STFC for his Rutherford Grant number ST/L003910/1.

\end{acknowledgements}

%
%

\end{document}